\documentclass{article}

\usepackage{arxiv}

\usepackage[utf8]{inputenc} % allow utf-8 input
\usepackage[T1]{fontenc}    % use 8-bit T1 fonts
\usepackage{hyperref}       % hyperlinks
\usepackage{url}            % simple URL typesetting
\usepackage{booktabs}       % professional-quality tables
\usepackage{amsfonts}       % blackboard math symbols
\usepackage{nicefrac}       % compact symbols for 1/2, etc.
\usepackage{microtype}      % microtypography
\usepackage{lipsum}
\usepackage{soul}
\usepackage{color}
\usepackage{float}
\usepackage{graphicx}
\usepackage{caption}
\usepackage{subcaption}
\usepackage{titlesec}
\usepackage{multirow}
\usepackage{amsmath}
\usepackage{soul,color}
\usepackage[section]{placeins}
\usepackage{array, makecell} %
\usepackage{float}
\usepackage[section]{placeins}
\usepackage{wrapfig}
\usepackage{enumitem}% http://ctan.org/pkg/enumitem
\usepackage{xspace,amstext}
\usepackage[table,xcdraw]{xcolor}

\usepackage[nodisplayskipstretch]{setspace}

\title{ Scalable deeper graph neural networks for high-performance materials property prediction}
\author{
Sadman Sadeed Omee\\
  Department of Computer Science and Engineering\\
  University of South Carolina\\
  Columbia, SC 29201 \\
  \And
Steph-Yves Louis\\
  Department of Computer Science and Engineering\\
  University of South Carolina\\
  Columbia, SC 29201 \\
  \And
 Nihang Fu, Lai Wei, Sourin Dey,  Rongzhi Dong, Qinyang Li\\
  Department of Computer Science and Engineering\\
  University of South Carolina\\
  Columbia, SC 29201 \\
  \And
 Jianjun Hu \thanks{S.Omee and S.Louis: equal contribution. Corresponding author: J.H. (http://www.cse.sc.edu/~jianjunh)}\\
  Department of Computer Science and Engineering\\
  University of South Carolina\\
  Columbia, SC, 29201, USA \\
  \texttt{jianjunh@cse.sc.edu} \\
}

\begin{document}
\maketitle

\begin{abstract}

Machine learning (ML) based materials discovery has emerged as one of the most promising approaches for breakthroughs in materials science. While heuristic knowledge based descriptors have been combined with ML algorithms to achieve good performance, the complexity of the physicochemical mechanisms makes it urgently needed to exploit representation learning from either compositions or structures for building highly effective materials machine learning models. Among these methods, the graph neural networks have shown the best performance by its capability to learn high-level features from crystal structures. However, all these models suffer from their inability to scale up the models due to the over-smoothing issue of their message-passing GNN architecture. Here we propose a novel graph attention neural network model DeeperGATGNN with differentiable group normalization and skip-connections, which allows to train very deep graph neural network models (e.g. 30 layers compared to 3-9 layers in previous works). Through systematic benchmark studies over six benchmark datasets for energy and band gap predictions, we show that our scalable DeeperGATGNN model needs little costly hyper-parameter tuning for different datasets and achieves the state-of-the-art prediction performances over five properties out of six with up to 10\% improvement. Our work shows that to deal with the high complexity of mapping the crystal materials structures to their properties, large-scale very deep graph neural networks are needed to achieve robust performances. 

\end{abstract}

\keywords{crystal structure prediction \and random crystal structure \and contact map \and differential evolution \and high symmetry }

\section{Introduction}

Machine learning models of materials properties have emerged as one of the most promising approaches for materials discovery due to their increasing prediction accuracy and their speed compared to the first principle calculations. Both composition and structure based ML models have been shown to be able to successfully predict materials properties, whose performance, however, is strongly dependent on the selection of the machine learning algorithm, the features, and the quality and amount of available datasets. Among these two types of screening models, the composition based ML models \cite{goodall2020predicting,wang2021compositionally} have the advantages of speed and capability to screen large-scale hypothetical compositions generated by generative deep learning models \cite{dan2020generative}. However, most or almost all materials properties are strongly dependent on the materials structures so that the structure based materials prediction models tend to have much higher prediction accuracy \cite{xie2018crystal,chen2019graph,dunn2020benchmarking}, which can be used to screen known materials structure repositories such as Inorganic Crystal Structure Database (ICSD)~\cite{bergerhoffinternational} or Materials Project Database~\cite{jain2013commentary}, or hypothetical crystal materials with structures created by modern generative deep learning models \cite{zhao2021high,nouira2018crystalgan}. Structural information of crystal materials can be represented in several methods \cite{liencoding} including  structure graph, Coulomb matrix~\cite{rupp2012fast}, topological descriptor, voronoi tessellation~\cite{chen2020critical}, diffraction fingerprint, or voxel grids \cite{zhao2020predicting}. However, due to the limited amount of structure data and very limited property labels, it remains an unsolved problem to achieve highly accurate materials property predictions from structures.

Currently, there are two major categories of ML approaches for structure based materials prediction based on their descriptors or features used: (1) the heuristic feature based models\cite{faber2015crystal,faber2016machine,ward2017including}, of which the features are designed based on existing physicochemical knowledge; (2) learned feature based models, of which the descriptors are learned by deep learning algorithms \cite{schutt2018schnet,xie2018crystal,chen2019graph}. While the heuristic feature based ML models have demonstrated some successes in a variety of applications such as formation energy prediction \cite{faber2015crystal}, ion conductivity screening \cite{sendek2017holistic}, large-scale benchmark studies have shown that the representation learning based deep graph neural network models have achieved much better performance in materials property prediction, which highlights the importance of developing more advanced deep learning models for materials property prediction \cite{fung2021benchmarking,rosen2021machine}. 

Since 2018, a variety of graph neural networks have been proposed to improve the prediction performance such as SchNet\cite{schutt2018schnet}, CGCNN\cite{xie2018crystal}, MEGNet \cite{chen2019graph}, MPNN\cite{gilmer2017neural}, iCGCNN \cite{park2020developing}, GATGNN\cite{louis2020global}, ALIGNN \cite{decost2021atomistic}. Each of these algorithms has utilized the graph representation as input along with slightly different additional information, convolution operators, and neural network architectures.  \cite{schutt2018schnet,xie2018crystal,chen2019graph}. However, a recent large-scale benchmark studies over five different datasets of varying sizes \cite{fung2021benchmarking} has shown that while the performance of existing graph neural network models are in general much better than those of non-GNN approaches, the performances of the best four GNN models tend to be saturated without significant difference. For example, for the Pt cluster dataset, the mean absolute error (MAE) values range from 0.151 (SchNet) to 0.205 (CGCNN). For the 2D materials dataset, the MAE range is 0.208 (CGCNN) to 0.224 (MEGNet). For the MOF dataset, the MAE range is between 0.228 (SchNet) and 0.253 (MEGNet). For the Alloy surface dataset, the MAE range is 0.058 (MPNN) to 0.069 (MEGNet). For the bulk crystal formation energy prediction problem, the MAE range is 0.046 (MPNN) and 0.05 (SchNet). Also, there are no dominant winners among these four GNN algorithms. After close investigation, we find that the optimized architectures of SchNet, MPNN, and CGCNN from the benchmark study have been set as (1-4-1 architecture) with 1 fully-connected layer plus 4 graph convolution layers plus an additional fully connected layer. For MEGNet, the architecture is 1-4-1-3 with 1 fully connected layer plus 4 graph convolution layers followed by 1 graph-convolution-fc layer and 1 fully connected layer, which also leads to much more weights compared to the other three GNN models. Overall, we find that all these models have only 4 or 5 graph convolution layers, which is in sharp contrast to those power deep learning models in computer vision and natural language processing where large deep neural networks dominates. The main obstacle that GNNs cannot go deeper is that most graph neural network models suffer from the over-smoothing issues \cite{cai2020note} in which all the node representations tend to become similar to each other with the increasing number of layers and makes these GNN models to be not scalable in terms of layers.

Inspired by the fact that large scale deep neural network models have led to breakthroughs in a variety of application domains. In computer vision, ResNet \cite{he2016deep,szegedy2017inception} and DenseNet~\cite{huang2017densely} with up to 1000 layers have been trained. In natural language modeling, the smallest GPT-3 model (125M parameters) has 12 attention layers and the largest GPT-3 model (175B parameters) uses 96 attention layers \cite{brown2020language}. Considering that in the materials property prediction problem, the number of element types and their sophisticated interactions are both much more complex than the pixels and their neighboring patterns, we expect very deep graph neural networks are needed to achieve significantly better results than the current state-of-the-art (SOTA) results as reported in the most recent benchmark study \cite{fung2021benchmarking}. Recently, Jha et al. \cite{jha2021enabling} applied the residual skip connection idea to vector input based materials property prediction. Their experiments showed that when the data set size is more than 15000, their individual residual networks outperform both the plain multi-layer perceptron networks and the stacked residual network. For only composition datasets with 234,299 samples, their 48-layer IRNet beat their 17-layer IRNet which outperform all other machine learning and plain neural network models. With vector represented structural features combined with composition features, their 17-layer IRNet achieve the best performance with up to 47\% performance improvement. However, no studies have been shown on graph neural networks which can better capture how the structural features affect their properties. Another study by Yang et al. \cite{yang2020learning} applied a deep convolution network with residual skip connections for crystal plasticity prediction with good performance. However, their models are still limited to vector representations instead of the graph representations. To our knowledge, there is no study on whether deeper graph neural networks can significantly push the frontier or state-of-the-art (SOTA) performance in materials property prediction.

In this work, we propose a very deep graph attention neural network model for large-scale materials property prediction with differentiable group normalization and residual skip connections. Our neural architecture allows us to train very deep graph neural networks with e.g. 30 or 50 layers compared to the current practice of 4-9 graph convolution layers. Our extensive experiments on the six benchmark datasets showed that our super-charged DeeperGATGNN models have achieved the SOTA results over the five out of six benchmark datasets with significant performance improvements with MAE errors reduction by up to 10\%. The model also has an especially attractive property: no tedious expensive hyper-parameter tuning is needed: only a sufficient number of graph convolution layers needs to be set. It also has much less risk of overfiting when too many graph convolution layers are included.
% We have also trained the largest/deepest graph neural networks for the bulk materials formation energy prediction with the lowest MAE error. 
We also applied our scaling strategy (differential group normalization plus skip connections) to other four graph neural networks and have achieved significant performance improvements for a few datasets for some models. We call these algorithms as DeeperCGCNN, DeeperMEGNet, DeeperMPNN, and DeeperSchNet.

Our contribution in this paper can be summarized as follows:

\begin{itemize}

    \item We identify the major challenge and bottleneck for graph neural networks for materials property prediction and propose increasing the depth of the networks to overcome the barrier.
    \item We propose a novel global attention based graph neural network architecture with differentiable group normization and residual skip connection to achieve scalable training of very deep graph neural network models for materials property prediction. The simplicity of our model with almost hassle free hyper-parameter tuning capability and scalability without worring about overfitting makes it ideal for large-scale materials property prediction
    \item We evaluate our DeeperGATGNN algorithm on six public benchmark datasets and achieved the best performance on five out of six datasets with up to 10\% performance improvements over previous SOTA resutls. 
    \item We demonstrate that our strategy of enabling deeper GNN for materials property prediction can also be applied to other four GNNs and achieve improved performance for some datasets. 
\end{itemize}

\section{Methods}

%\subsection{Prediction of crystal structure based on contact map}
\subsection{Overall architecture for deep graph attention neural networks}
% \hl{summary of the basic idea  by Steph-yves}
Our method of deeper Graph Neural Network is based on our previously proposed graph neural network model GATGNN \cite{louis2020global}. In this model,we developed a graph neural network that uses two variants of graph soft-attention to learn properties of inorganic molecules \cite{velivckovic2017graph,wang2019heterogeneous,zhang2018gaan,liu2019geniepath,louis2021node}. The first type of soft-attention consists of additive multi-head attention (4 or 8) applied to the 1-hop neighbors of each atom. These attention layers are only used to extract the locally dependent features between neighboring atoms. Afterwards, upon extracting the local features, GATGNN then uses a unique soft-attention at the end to transform neighborhood dependent information to a global context (with respect to all other atoms in the crystal). The local soft-attention  $\alpha_{i,j}$ between a node $i$ and a neighbor $j$ can be represented by the as:
\begin{equation}\label{eq:1}
    \alpha_{i,j} = \frac{\text{exp}(a_{i,j})}{\sum_{k \in N_{i}}{\text{exp}(a_{i,k})}}
\end{equation}
In Eq.~\ref{eq:1}, $N_{i}$ represents the neighborhood of node $i$ and $a_{i,j}$ is the parameterized weight coefficient between nodes $i$ and $j$, which represents the importance of node $\_$ to node $i$. The global attention: $g_i$, which is applied right before the global pooling, calculates the overall importance of each node. It can be described as the following equation:
\begin{equation}\label{eq:2}
    g_{i} = \frac{ (\textbf{x}_{i}\parallel \textbf{E}) \cdot \textbf{W}} {\sum_{x_c \in \textbf{X}}{(\textbf{x}_c \parallel \textbf{E}) \cdot \textbf{W}}}
\end{equation}
In Eq.~\ref{eq:2}, $\textbf{x} \in \mathbb{R}^{F}$ represents a learned embedding, $\textbf{E}$ a compositional vector of the crystal, $\textbf{W} \in \mathbb{R}^{1\times (F+|E|)}$ a parameterized matrix, and $x_c$ is the learned embedding of any atom $c$ within the crystal. 
 By using the combination of 3 to 5 of these local soft-attention and one global soft-attention, GATGNN was able to match state-of-the-art of inorganic materials properties prediction for most of the properties (except formation energy) at the time and also provide interpretable results in terms of the contribution of each atom. Nevertheless, the issue of over-smoothing: a general challenge that prevents the use of more than a few layers in general graph neural networks also affects GATGNN. With the use of more than 7 layers, the performance of the GATGNN algorithm begins to considerably decrease with the addition of more layers. With the expectation that a deeper model should be able to extract even more of these inter-atomic dependent features, we aim to overcome this over-smoothing limitation so our model can more effectively extract the physics-dependent features of crystals. Our devised solution consists of using additive skip-connections between these attention layers that extract the local features and further improve the learning with the addition of differentiable normalization layers.

\begin{figure}[ht]
  \centering
  \includegraphics[width=\linewidth]{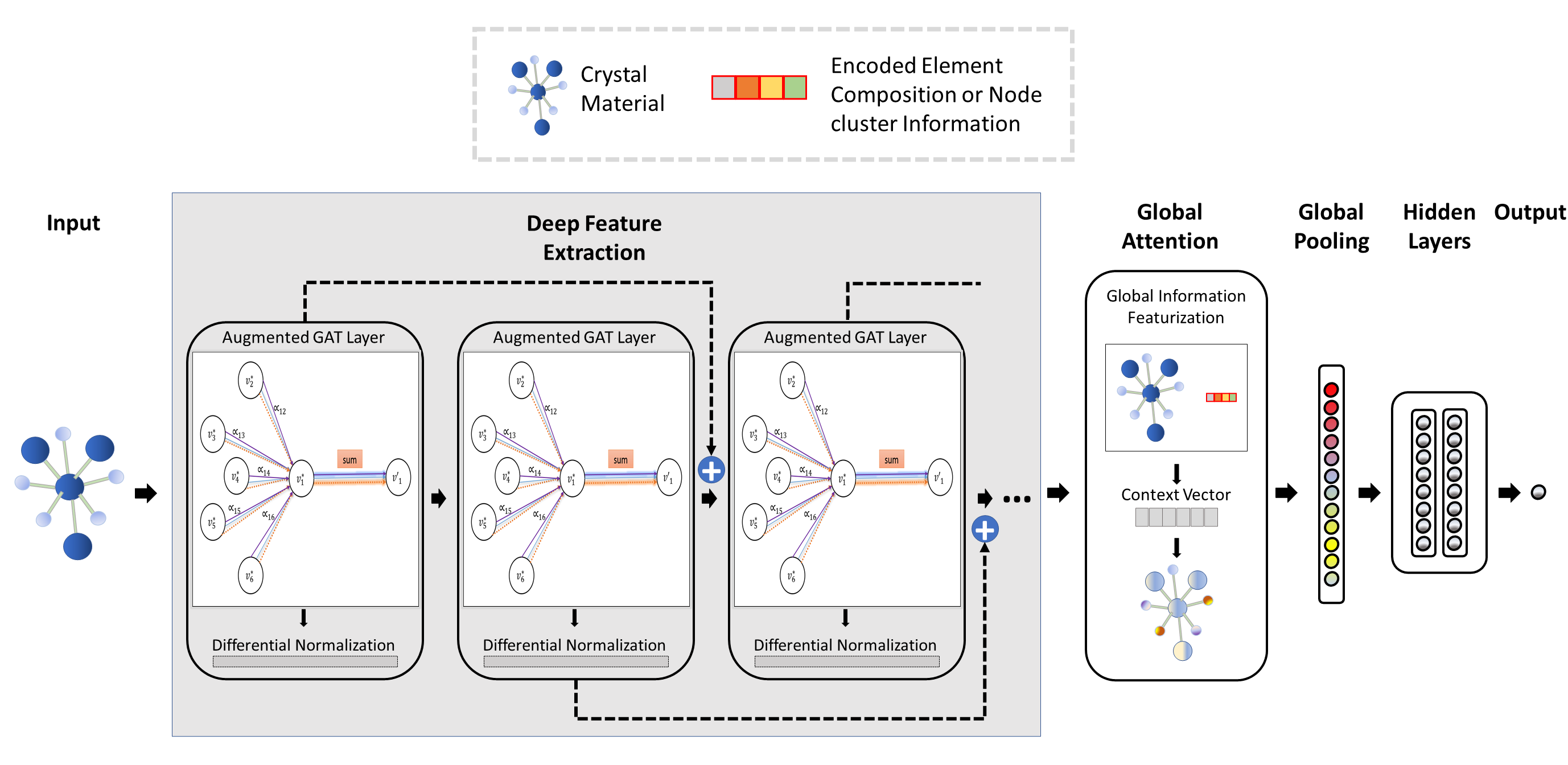}
  \caption{Architecture of the scalable global attention graph neural network.}
  \label{fig:framework}
\end{figure}

 The whole architecture of the proposed DeeperGATGNN model is shown in Figure \ref{fig:framework}. It consists of several augmented GAT attention layers with skip connections and differentiable normalization operators, which is followed by a global attention layer and global pooling layer. Finally a few fully-connected layers are added before the output layer.

\subsection{Differentiable Group Normalization}

One of the major issues in training a deep GNN architecture is the over-smoothing problem, in which  the representation vectors of all nodes of a graph become indistinguishable as the number of graph convolution layers increases~\cite{li2018deeper,oono2019graph,chen2020measuring,luan2019break,chen2020simple,zhou2020graph}. This problem restricts GNNs to only a very few layers for better performance~\cite{kipf2016semi,velivckovic2017graph,wu2020comprehensive}. For example, both GAT~\cite{velivckovic2017graph} and GCN~\cite{kipf2016semi} perform best when the number of layers is limited to only 2 and thus they fail to utilize higher order neighbors' features~\cite{chen2020simple}. Many approaches have already been adopted to improve upon the problem~\cite{rong2019dropedge,xu2018representation,li2019deepgcns,zhao2019pairnorm,zhou2020towards}. Traditional normalization techniques used to reduce this problem include Batch Normalization~\cite{ioffe2015batch} or measures based on node pair distances such as Pair Normalization~\cite{zhao2019pairnorm}. But these techniques do not take account of the global graph structure which results in sub-optimal performance when the number of layers of the GNN is large~\cite{zhou2020towards}. Recently, Zhou et al.~\cite{zhou2020towards} addressed this issue by proposing the Differentiable Group Normalization (DGN). The main procedure of DGN is to first use a cluster assignment matrix to cluster the nodes of a graph to different clusters and then normalize each cluster separately. This will make the representations of nodes within the same community/class to be similar while those of different classes to be separated, leading to effective control of the over-smoothing issue. Due to their simplicity of implementation, we have adopted the DGN normalization in our DeeperGATGNN architecture. More specifically, the DGN works as follows:

Let, $n$ be the number of nodes, $G$ be the number of clusters specified. Let, $H^{(l)} \in \mathbb{R}^{n\times d^{(l)}}$ be the embedding matrix derived after the $l$-th layer of a GNN where $d(l)$ is the embedding dimension of layer $l$. Then the cluster assignment matrix $S^{(l)}$ can be calculated using the following equation where $U^{(l)} \in \mathbb{R}^{d^{(l)}\times G}$ is a trainable parameter:

\begin{equation}
    S^{(l)} = softmax(H^{(l)}U^{(l)})
\end{equation}

The cluster assignment matrix $S^{(l)} \in \mathbb{R}^{n\times G}$ stores the probabilities of each node of the graph of being assigned to each cluster. It then places the nodes into different groups using the following equation:

\begin{equation}\label{eq:nodes_assign}
    H_{i}^{(l)} = S^{(l)}[:,i] \circ H^{(l)}\\
\end{equation}

$H_{i}^{(l)}$ denotes the embedding matrix for cluster $i$ and $\circ$ is the row-wise multiplication operator in Eq.~\ref{eq:nodes_assign}. Each cluster is then normalized separately using the following equation:

\begin{equation}\label{eq:separate_norm}
    \Tilde{H}_{i}^{(l)} = \gamma_i(\frac{H_{i}^{(l)} - \mu_i}{\sigma_i}) + \beta_i
\end{equation}

$\mu_i$ and $\sigma_i$ means the mean and standard deviation of each group $i$ in Eq.~\ref{eq:separate_norm} and $\gamma_i$ and $\beta_i$ denotes two trainable parameters.

Finally, the final embedding matrix can be calculated using the following equation:

\begin{equation}\label{eq:final_embedding}
    \Tilde{H}^{(l)} = H^{(l)} + \lambda\sum_{i=1}^G \Tilde{H}_{i}^{(l)} \in \mathbb{R}^{n\times d^{(l)}}
\end{equation}

$\lambda$ denotes a balancing factor and $ \Tilde{H}^{(l)} = [ \Tilde{H}_1^{(l)}, \Tilde{H}_2^{(l)}, \ldots, \Tilde{H}_G^{(l)}]$ denotes the final embedding matrix in Eq.~\ref{eq:final_embedding}.

The two main reasons why DGN is so successful in preventing the over-smoothing issue are: (1) each group is normalized separately using Eq.~\ref{eq:separate_norm}, so each group will have different mean and standard deviation and thus the probability of the representation vectors of nodes of different groups being similar will decrease and (2) input embedding is preserved in Eq.~\ref{eq:final_embedding} to prevent over-normalization.

\subsection{Skip connections for enabling scalable graph neural networks}

One of the major enabling techniques in deep learning for training very deep networks is the residual skip connection as first introduced in the ResNet framework \cite{he2016deep}, which allows it to train networks with up to 1000 layers. The key idea of residual skip connection is to learn to achieve identity mapping where the input x is added to the output of stacked layers so we have H(x)=F(x)+x. So, instead of learning the underlying mapping H(x) function, the stacked layers  are used to learn the residual mapping F(x) = H(x)- x. The major benefit is that if the identity mapping is already optimal and the stacked layers cannot learn more salient information, the training procedure can push the residual mapping to zero so as to avoid the degradation problem. So compared to conventional multi-layer networks where a stack of layers are trained to directly learn the desired underlying mapping, the  layers in a residual module learns a residual mapping. The residual skip connection idea can be applied to stack any kind of network layers such as convolutional or fully connected layers or graph convolution layers. Not surprisingly, residual connections have been introduced into graph neural networks for training deeper networks \cite{li2020deepergcn,li2021training,xu2021optimization,chi2021residual}. Concepts such as residual/dense connections and dilated convolutions have also been transferred from CNNs to GCNs in order to successfully train very deep GCNs \cite{li2021deepgcns}. 

In this work, we implemented the layer-wise residual skip connections in our global attention graph neural networks, which is similar to the IRNet \cite{jha2021enabling}.

\begin{figure}[H] 
\centering
\begin{minipage}{0.8\linewidth}
    \begin{subfigure}[t]{0.4\textwidth}
        \includegraphics[width=\textwidth]{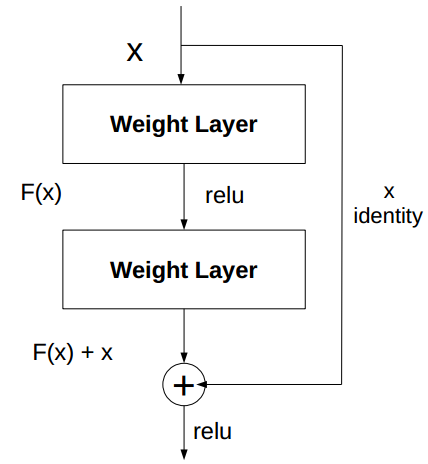}
        \caption{Skip connection in ResNet with 2-layer skips}
        \vspace{-3pt}
        \label{fig:resnet}
    \end{subfigure}\hfill
    \begin{subfigure}[t]{0.3\textwidth}
        \includegraphics[width=\textwidth]{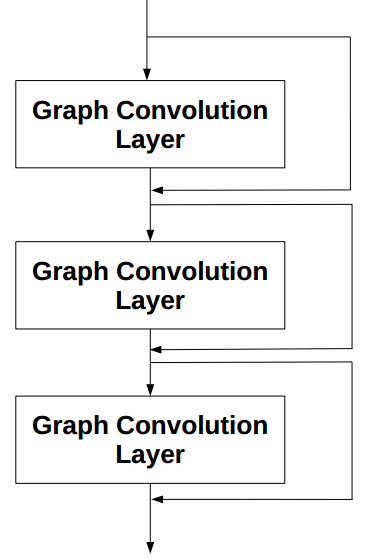}
        \caption{Skip connection in DeeperGATGNN}
        \vspace{-3pt}
        \label{fig:La3Al1N1}
    \end{subfigure}
    
    \caption{Comparison of the skip connection in ResNet and in DeeperGATGNN}
    \label{fig:skipconnection}
\end{minipage}
\end{figure}

\subsection{Evaluation criteria}

To evaluate the performance of our models, we use mean absolute error (MAE), mean square error(MSE) and $R^2$ (coefficient of determination) which are all standard evaluation criteria for regression problems. We use both 5-fold cross-validation and hold-out tests for performance evaluations depending on specific experiments. The parameters of the base line models are specified by the supplementary file in \cite{fung2021benchmarking}. Our GATGNN and DeeperGATGNN model hyper-parameters are provided in the supplementary file.

\section{Experimental Results}
\label{sec:headings}

\subsection{Datasets}

We evaluate the performances of our models and other baselines using six datasets including five benchmark datasets used in previous evaluation study \cite{fung2021benchmarking}. The first five datasets are all for formation/surface energy predictions of nanoclusters materials (pt-cluster), alloy surface, bulk materials downloaded from MP database, 2D materials, and MOF materials. We added an additional band gap dataset for the same set of bulk materials. The details of the datasets are shown in Table \ref{table:dataset}. The number of samples ranges from 3814 to 37334 and the number of elements from 1 to 87, reflecting the diversity of the datasets and the challenges in predicting corresponding materials properties.

\begin{table}[htb!] 
\begin{center}
\caption{Six benchmark datasets}
\label{table:dataset}
\begin{tabular}{|l|l|l|l|l|l|l|l|l|l|}
\hline
\textbf{\makecell{Dataset}} & 
\textbf{\makecell{source}} & 
\textbf{\makecell{\# of elements}} &
\multicolumn{1}{c|}{\textbf{\# of samples}} \\ \hline

Bulk materials formation energy              & MaterialsProject \cite{jain2013commentary}                 & 87        & 36839              \\ \hline
Alloy surface energy          & \cite{dai2018syntax}    & 42               & 37334           \\ \hline
Pt-cluster formation energy          & \cite{dai2018syntax}    & 1               & 19801           \\ \hline
2D materials formation energy             & \cite{haastrup2018computational}                 & 60        & 3814    \\ \hline
MOF formation energy            & \cite{rosen2021machine}                & 4        & 18321        \\ \hline

Bulk materials band gap             & MaterialsProject  \cite{jain2013commentary}                 & 87        & 36837            \\ \hline

\end{tabular}
\end{center}
\end{table}

\FloatBarrier

\subsection{Performance Comparison of DeeperGATGNN with other state-of-the-art GNN models}
\label{subsec: comparison}

To fairly and objectively evaluate and compare the performance of our models with other state-of-the-art models, it is critical to ensure all models are trained on the same datasets with appropriate training and optimal hyper-parameters and using the same cross-validation method. An excellent largest benchmark study by Fung et al.\cite{fung2021benchmarking} implemented seven different prediction models (including four graph neural network models) and a dummy baseline model in the same code base and evaluated their performance on the same set of five datasets using large-scale computationally expensive hyper-parameter tuning (limited to 200 epochs due to computational burden) to identify eight optimal parameters for all models on different datasets. To compare, we re-implemented our algorithm in their code framework and use a single parameter setting (with 10,15,20,25 graph convolution layers) for all the five datasets plus an additional band gap dataset (See supplementary file for hyper-parameter setting). All the models are trained with 2500 epochs. The results are shown in Table \ref{tab:sota-performance}, in which the results for SchNet, MPNN, CGCNN, and MEGNet are all from previous benchmark study \cite{fung2021benchmarking}.

First, we find that the benchmark study \cite{fung2021benchmarking} shows that for different datasets, there are different winning algorithms with very different hyper-parameter settings except that MEGNet does not win on any of the six datasets. Our previous GATGNN achieves better result on the bulk crystal formation energy prediction problem than other four models (SchNet, MPNN, CGCNN, and MEGNet). However, with our DeeperGATGNN with 20 layers, it beats all the previous best results. We further tried 25 and 30 graph convolution layers, which leads to further improvements for 2D materials dataset and band gap problem. The last line of Table~\ref{tab:sota-performance} summarizes the percentage of performance improvements ranging from 5.34\% (for band gap prediction) to 34.97\% (for bulk formation energy prediction), all achieved with a single hyper-parameters across 6 different datasets. Compared to our previous GATGNN model, by addressing the over-smoothing issue using the differentiable normalization and skip connection, our DeeperGATGNN models achieves from 10.07\% to 54.04\% reduction in the MAE prediction errors across the six materials prediction problems.

\begin{table}[]
\centering
\caption{Comparison of 5-fold cross-validation performance of DeeperGATGNN to benchmark results.}
\label{tab:sota-performance}
\begin{tabular}{|l|c|c|r|r|r|r|r|}
\hline
ML models                             & GC-layer              & \begin{tabular}[c]{@{}c@{}}Bulk crystals\\ (Eform) \end{tabular}                           & \multicolumn{1}{c|}{\begin{tabular}[c]{@{}c@{}}Alloy\\surfaces \end{tabular} }          & \multicolumn{1}{c|}{MOFs}                    & \multicolumn{1}{c|}{\begin{tabular}[c]{@{}c@{}}2D\\ Materials \end{tabular} }            & \multicolumn{1}{c|}{Pt clusters}             & \multicolumn{1}{c|}{\begin{tabular}[c]{@{}c@{}}Band gap\\ (bulk) \end{tabular}}            \\ \hline
\multicolumn{1}{|r|}{Dataset size}    &                       & 36839                                          & 37334                                        & 18321                                        & 3814                                         & 19801                                        & 36837                                          \\ \hline
SchNet                                & Misc                  & 0.05                                           & 0.063                                        & \cellcolor[HTML]{FFCC67}{\ul \textit{0.228}} & 0.214                                        & \cellcolor[HTML]{FFC702}{\ul \textit{0.151}} & 0.28168                                        \\ \hline
MPNN                                  & Misc                  & 0.046                                          & \cellcolor[HTML]{FFC702}{\ul \textit{0.058}} & 0.245                                        & \cellcolor[HTML]{FFCB2F}{\ul \textit{0.204}} & 0.182                                        & 0.26485                                        \\ \hline
CGCNN                                 & Misc                  & 0.049                                          & 0.060                                         & 0.233                                        & 0.208                                        & 0.205                                        & \cellcolor[HTML]{FFC702}{\ul \textit{0.25977}} \\ \hline
MEGNet                                & Misc                  & 0.048                                          & 0.069                                        & 0.253                                        & 0.224                                        & 0.180                                         & 0.26485                                        \\ \hline
GATGNN                                & 5                  & \cellcolor[HTML]{FFCC67}{\ul \textit{0.04544}} & 0.08063                                      & 0.24222                                      & 0.20745                                      & 0.28249                                      & 0.27341                                        \\ \hline
DeeperGATGNN                          & 10                    & 0.03019                                        & 0.05022                                      & 0.22384                                      & 0.19161                                      & \textbf{0.12982}                             & 0.26243                                        \\ \hline
DeeperGATGNN                          & 15           & \textbf{0.02955}                               & 0.04158                                      & 0.21775                                      & 0.21387                                      & 0.13629                                      & 0.25586                                        \\ \hline
DeeperGATGNN                          & 20                    & 0.02968                                        & \textbf{0.04086}                             & \textbf{0.21583}                             & 0.17745                                      & 0.13210                                       & 0.25504                                        \\ \hline
DeeperGATGNN                          & 25                    & 0.03056                                        & 0.04112                                      & 0.21688                                      & \textbf{0.17185}                             & 0.14126                                      & \textbf{0.24570}                                \\ \hline
DeeperGATGNN                          & 30                    & 0.03041                                        & 0.04268                                      & 0.21782                                      & 0.17300                                        & 0.15218                                      & 0.24594                                        \\ \hline
\begin{tabular}[c]{@{}c@{}}\% of improvements\\ over previous best\end{tabular}   & \multicolumn{1}{l|}{} & \multicolumn{1}{c|}{34.97\%}                      & \multicolumn{1}{c|}{29.55\%}                    & \multicolumn{1}{c|}{5.34\%}                     & \multicolumn{1}{c|}{15.76\%}                    & \multicolumn{1}{c|}{14.03\%}                    & \multicolumn{1}{c|}{5.42\%}                       \\ \hline
\begin{tabular}[c]{@{}c@{}}\% of improvements\\ over GATGNN\end{tabular}   & \multicolumn{1}{l|}{} & \multicolumn{1}{c|}{34.97\%}                      & \multicolumn{1}{c|}{49.32\%}                    & \multicolumn{1}{c|}{10.07\%}                     & \multicolumn{1}{c|}{17.16\%}                    & \multicolumn{1}{c|}{54.04\%}                    & \multicolumn{1}{c|}{10.13\%}                       \\ \hline
\end{tabular}
\end{table}

Next, we wonder if the number of epochs has made the performance difference. So we plot the training/validation errors over the training process for all the six GNN models, two of them are shown in Figure \ref{fig:stagant} and others are shown in Supplementary file. We found that with the current hyper-parameter settings (especially the learning rate scheduling), almost all models stagnate around 500 rather than 200 epochs which are used in previous benchmark study \cite{fung2021benchmarking}. It seems their reported results for different models may be under-estimated due to insufficient training. To validate this, we re-train all their models using 500 epochs while keeping all other optimal hyper-parameters they acquired unchanged. The new results and the best results of our models are summarized in Table \ref{tab:newresults}.

We find that the optimal performances of almost all algorithms have been significantly improved by increasing the training epochs to 500. For example, for Bulk crystal dataset, the best performance is now achieved by MEGNet with MAE of 0.03295 eV a significant 31.37\% improvement from the 0.048 eV as reported in \cite{fung2021benchmarking}. For Alloy surface dataset, the best model is CGCNN with an MAE of 0.04239 with 29.35\% improvement from previous result. Overall, for previous generation of models, the best one is MEGNet with the best results on three datasets, followed by CGCNN with the best performance on two datasets. MPNN however achieves superior result on the Pt-cluster dataset. However, our DeeperGATGNN models trained with the single hyper-parameter settings (except the graph convolution layers, which is easy to test) achieved the best results over five out of six datasets, with performance improvements from 0.05\% to 10.29\%. Since training such large-scale graph neural networks are very computationally intensive (some models take 1 to 2 days to train), finding optimal hyper-parameters with large epochs is infeasible. In this case, the easy hyper-parameter setting of our DeeperGATGNN is a very attractive feature combined with its outstanding state-of-the-result performance. This can be found by the variety of the graph convolution (GC) layers used in the optimal models of previous algorithms. For example, the MEGNet uses 4 to 8 GC layers across the six datasets while CGCNN uses 1 to 8 GC layers. The MPNN models use 2 to 5 GC layers while SchNet uses 1 to 9 GC layers. In our case, a single setting of 20 or 25 GC layers allows our model to achieve great results.

\begin{figure}[ht]
\centering
    \begin{subfigure}[h]{0.4\textwidth}
        \includegraphics[width=\textwidth]{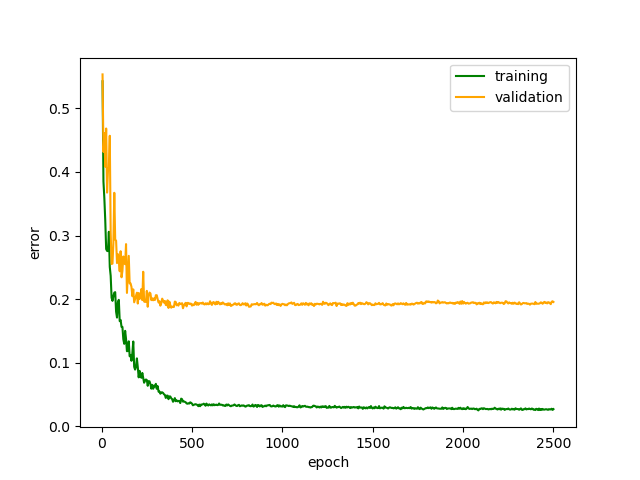}
        \caption{MEGNet}
        \vspace{-3pt}
        \label{fig:all_fe_violin}
    \end{subfigure}
    \begin{subfigure}[h]{0.4\textwidth}
         \includegraphics[width=\textwidth]{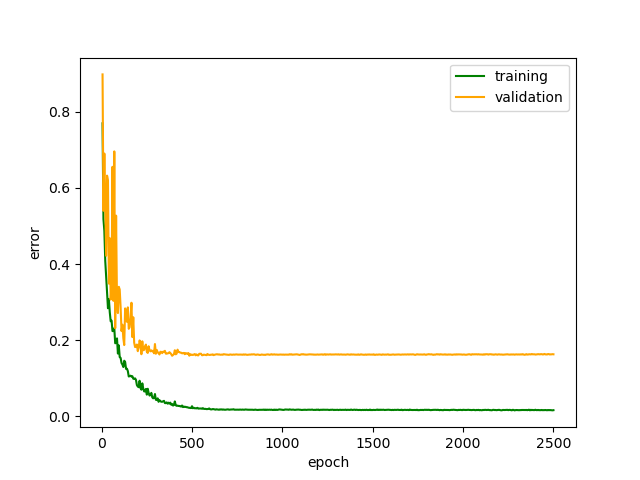}
        \caption{DeeperGATGNN}
        \label{fig:low_fe_violin}
    \end{subfigure}
    
 \caption{Training/validation errors of MEGNet and our DeeperGATGNN. All models become stagnant after 500 epochs. More epochs do not have chance to further improve their performances. So our GATGNN models trained with 2500 epochs should have similar performance when trained with 500 epochs.}
    \label{fig:stagant}
\end{figure}

% \usepackage[normalem]{ulem}
\begin{table}[ht]
\centering
\caption{Comparison of DeeperGATGNN performance with corrected benchmark results. (5-fold CV MAE. gcl means the number of graph convolution layers of the corresponding optimal models)}
\label{tab:newresults}
\begin{tabular}{|l|r|r|r|r|r|r|r|r|r|r|r|r|}
\hline
ML models      & \multicolumn{1}{c|}{\begin{tabular}[c]{@{}c@{}}Bulk\\crystal \end{tabular}}      & \multicolumn{1}{l|}{gcl} & \multicolumn{1}{l|}{\begin{tabular}[c]{@{}c@{}}Alloy\\surfaces \end{tabular}}   & \multicolumn{1}{l|}{gcl} & \multicolumn{1}{l|}{MOFs}             & \multicolumn{1}{l|}{gcl} & \multicolumn{1}{l|}{\begin{tabular}[c]{@{}c@{}}2D\\materials \end{tabular}}     & \multicolumn{1}{l|}{gcl} & \multicolumn{1}{l|}{\begin{tabular}[c]{@{}c@{}}Pt\\clusters \end{tabular}}      & \multicolumn{1}{l|}{gcl} & \multicolumn{1}{l|}{\begin{tabular}[c]{@{}c@{}}Band\\gap \end{tabular}}   & \multicolumn{1}{l|}{gcl}   \\ \hline
SchNet         & 0.05561                                        & 1                        & 0.04696                               & 5                        & 0.23124                               & 9                        & 0.22402                               & 4                        & 0.17257                               & 9                        & 0.28168                               & 4                                               \\ \hline
MPNN           & 0.03485                                        & 5                        & 0.04879                               & 5                        & 0.20701                               & 4                        & 0.18929                               & 2                        & \cellcolor[HTML]{FFD966}{ 0.13842} & 3                        & 0.26485                               & 4                                                 \\ \hline
CGCNN          & 0.03492                                        & 7                        & \cellcolor[HTML]{FFD966}{ 0.04239} & 8                        & 0.2141                                & 6                        & 0.20009                               & 6                        & 0.30242                               & 1                        & \cellcolor[HTML]{FFD966}{\ul 0.25977} & 4                                                \\ \hline
MEGNet         & \cellcolor[HTML]{FFD966}{ 0.03294} & 5                        & 0.04678                               & 8                        & \cellcolor[HTML]{FFD966}{ 0.19682} & 7                        & \cellcolor[HTML]{FFD966}{ 0.17194} & 4                        & 0.28766                               & 8                        & 0.26485                               & 4                                                 \\ \hline
GATGNN         & 0.04752                                        & 5                        & 0.07265                               & 5                        & 0.2265                                & 5                        & 0.18687                               & 5                        & 0.17476                               & 5                        & 0.27341                               & 5                                                 \\ \hline
Ours   & \textbf{0.02955}          & 15                       & \textbf{0.04086}                      & 20                       & 0.21583                               & 20                       & \textbf{0.17185}                      & 25                       & \textbf{0.12982}                      & 10                       & \textbf{0.2457}                       & 25                                               \\ \hline
\% improve & \textbf{10.29\%}                               & \multicolumn{1}{l|}{}    & \textbf{3.61\%}                       & \multicolumn{1}{l|}{}    & -9.66\%                               & \multicolumn{1}{l|}{}    & \textbf{0.05\%}                       & \multicolumn{1}{l|}{}    & \textbf{6.21\%}                       & \multicolumn{1}{l|}{}    & \textbf{5.42\%}                       & \multicolumn{1}{l|}{}        \\ \hline
\end{tabular}
\end{table}

We further check whether increasing the number of GC layers can improve the performance of existing GNN models. The answer is no as is shown in our scalability study (Figure~\ref{fig:scalability}). Increasing the GC layers moderately actually deteriorate the performance for SchNET and GATGNN and also for CGCNN and MEGNet with a lower degree. However, when 30 GC layers are used, all the other algorithms including the Deeper version of the existing models totally collapse except DeeperGATGNN. We want to note that despite that our DeeperGATGNN model has much more layers than other baselines, the number of parameters of our model is actually similar. For example the MEGNet model with 6/8 layers has 1.2/1.6 million parameters while our DeeperGATGNN with 20/25 GC layers has only 0.9/1.1 million parameters respectively, indicating the high parsimony of our model.

\begin{figure}[ht]
\centering
    \begin{subfigure}[h]{0.8\textwidth}
        \includegraphics[width=\textwidth]{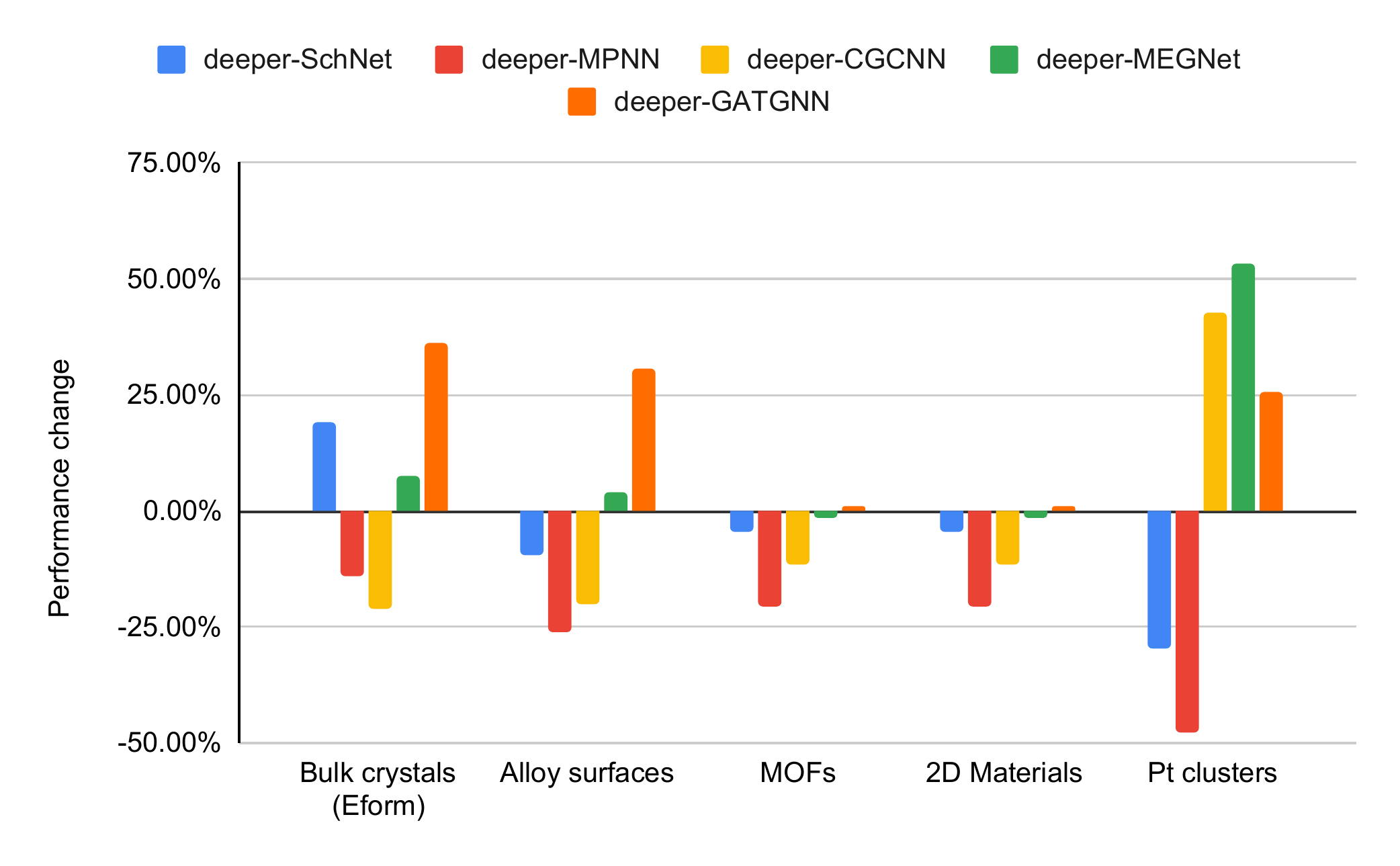}
    \end{subfigure}

 \caption{Performance change via using differentiable normalization and skip connections.}
    \label{fig:performancechange}
\end{figure}

As most existing message passing based GNN models suffer from the over-smoothing issue, we check if the differentiable normalization and skip connection can help improve other GNN models so that they can also benefit from deeper GC layers. To verify this, we increase the GC layers to 10 for all the models as evaluated in the benchmark study, replace their original weight normalization method with differentiable normalization and add the skip connections as shown in Figure \ref{fig:skipconnection}, and train the models using their optimal hyper-parameters except that we train with 500 epochs. Then we calculate the performance changes after these modifications. The results are shown in Figure \ref{fig:performancechange}. First, for the SchNet model, the DeeperSchNet achieves 19\% reduction in MAE error for the bulk dataset while getting worse results for all other datasets. For the Pt clusters, its performance dropped 29.5\%. For the MPNN model, all enhanced models achieve much worse results ranging from -14.2\% to -47.7\%, demonstrating its lack of scalability. For CGCNN, surprisingly for all datasets except the Pt clusters, its performance all becomes worse. However, for the Pt clusters, the DeeperCGCNN reduces the MAE error of CGCNN by almost 43\% from 0.30 to 0.17 eV. MEGNet and GATGNN are the only two models that all get boosting from adding differentiable normalization and skip connection. But the performance improvements of GATGNN are much higher, especially for the bulk crystal, alloy  surfaces, and the Pt clusters.

\begin{figure}[ht]
\centering
    \begin{subfigure}[h]{0.95\textwidth}
        \includegraphics[width=\textwidth]{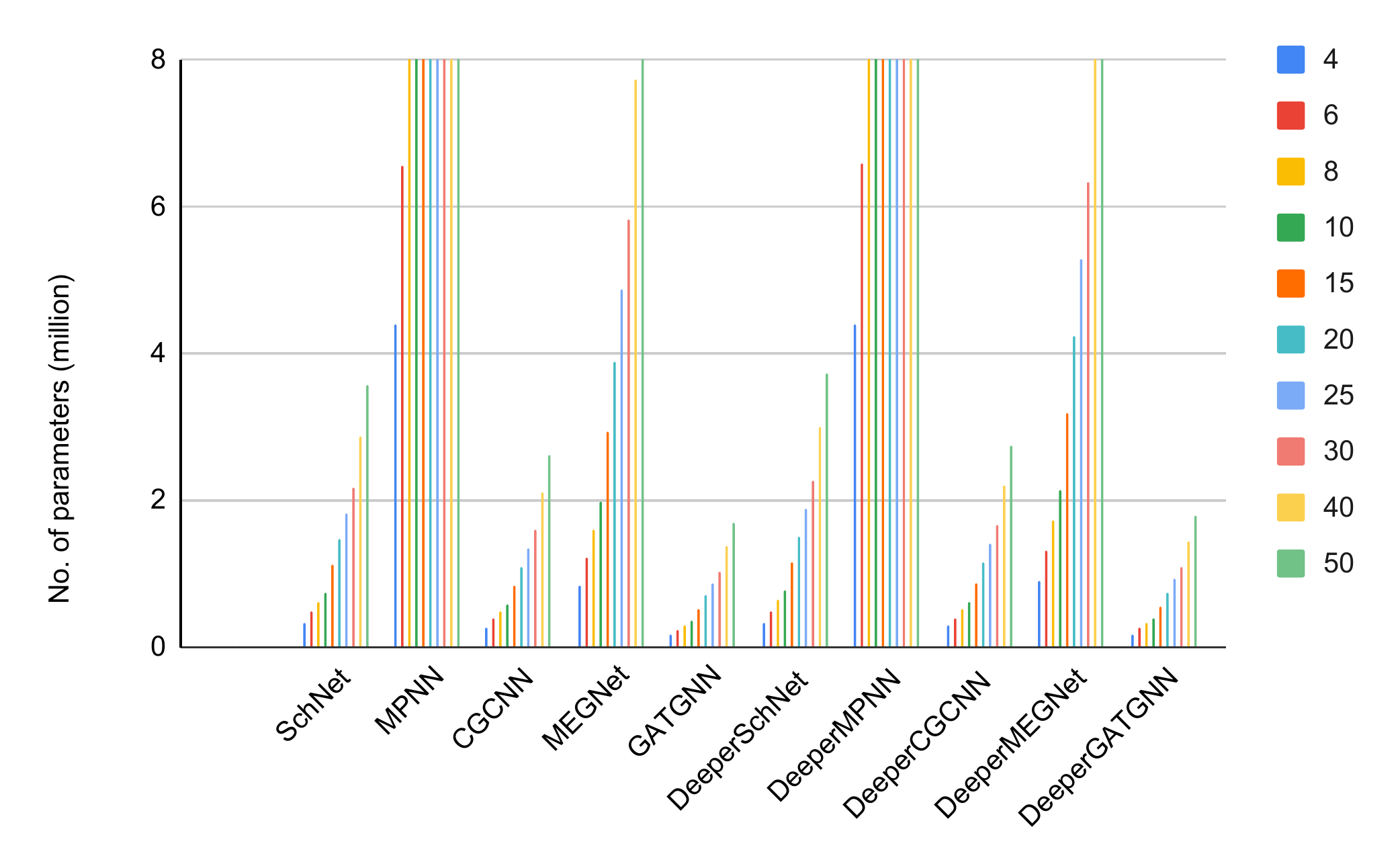}
    \end{subfigure}

 \caption{Parameter numbers for GNN models with increasing number of graph convolution layers.}
    \label{fig:parameterno}
\end{figure}

To further explain why the existing GNN models benefit little from the differentiable normalization and skip connections in most cases, we plot the number of parameters for these models and their deeper versions as shown in Figure \ref{fig:parameterno}. First we find that for MPNN and DeeperMPNN, the number of parameters increase very rapidly reaching to more than 8 millions when the number of GC layers reach 8. With limited training samples, it does cause some serious problem. The second most parameter-rich models are MEGNet and DeeperMEGNet, both have almost 6 million parameters when the number of GC layers approximate 30, which lead to their collapsed performance as shown in our scalability study in Figure \ref{fig:scalability}. The CGCNN and SchNet models along with their variants are more parameter-parsimonious, but adding more layers do not improve the results except for some special datasets. Our GATGNN and DeeperGATGNN are the most parsimonious models. Even with 50 GC layers, the number of parameters is under 1.8 million compared to MEGNet and DeeperMEGNet's more than 9 million parameters, which may partially expain why DeeperGATGNN can improve the base model significantly.

\begin{figure}[ht!] 
\centering
\begin{minipage}{1.0\linewidth}

    \begin{subfigure}[t]{0.33\textwidth}
        \includegraphics[width=\textwidth]{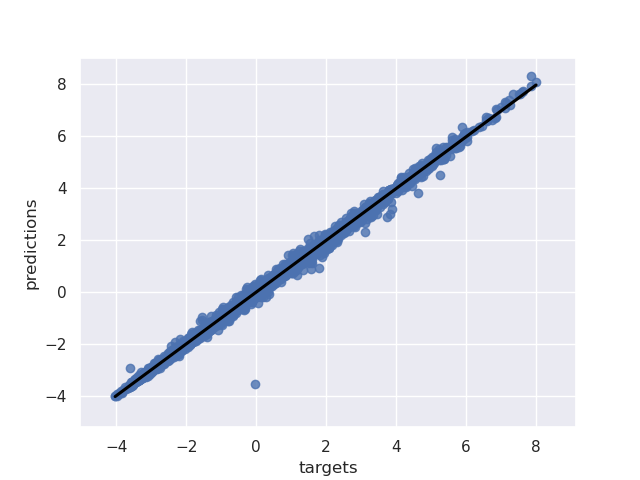}
        \caption{ SchNet (MAE: 0.047 eV) }
        \vspace{-3pt}
        \label{fig:spSc2D}
    \end{subfigure}
    \begin{subfigure}[t]{0.33\textwidth}
        \includegraphics[width=\textwidth]{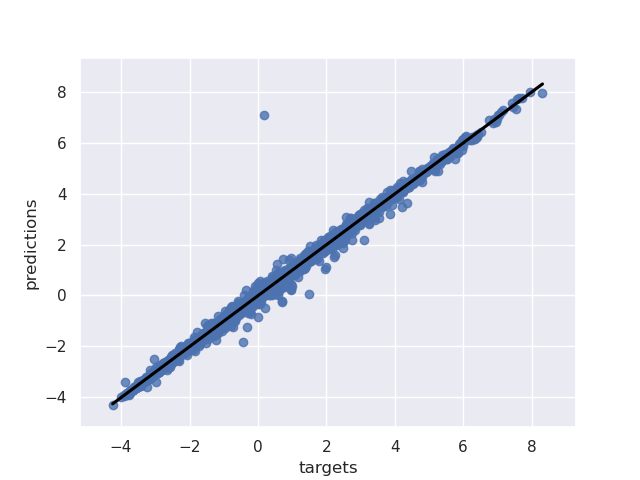}
        \caption{ MPNN (MAE: 0.0488 eV)}
        \vspace{-3pt}
        \label{fig:spMp2D}
    \end{subfigure}
    \begin{subfigure}[t]{0.33\textwidth}
        \includegraphics[width=\textwidth]{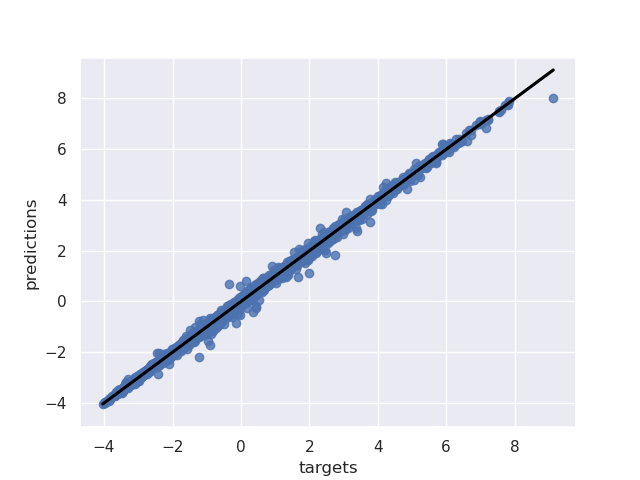}
        \caption{ CGCNN (MAE: 0.0424 eV)}
        \vspace{-3pt}
        \label{fig:spCg2D}
    \end{subfigure}\hfill
    
    \begin{subfigure}[t]{0.33\textwidth}
        \includegraphics[width=\textwidth]{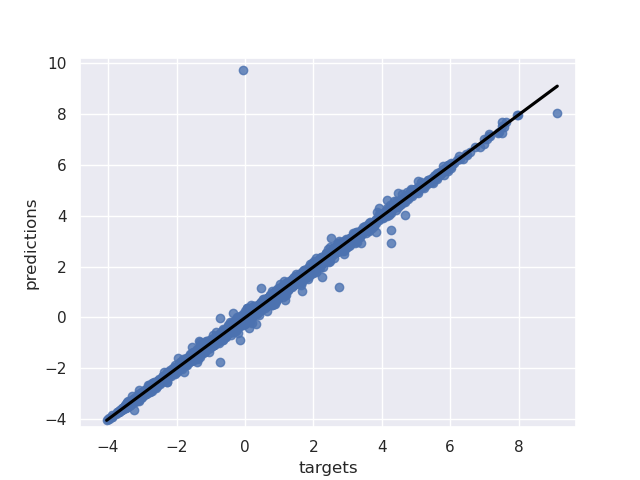}
        \caption{ MEGNet (MAE: 0.0468 eV)}
        \vspace{-3pt}
        \label{fig:spMe2D}
    \end{subfigure}
    \begin{subfigure}[t]{0.32\textwidth}
        \includegraphics[width=\textwidth]{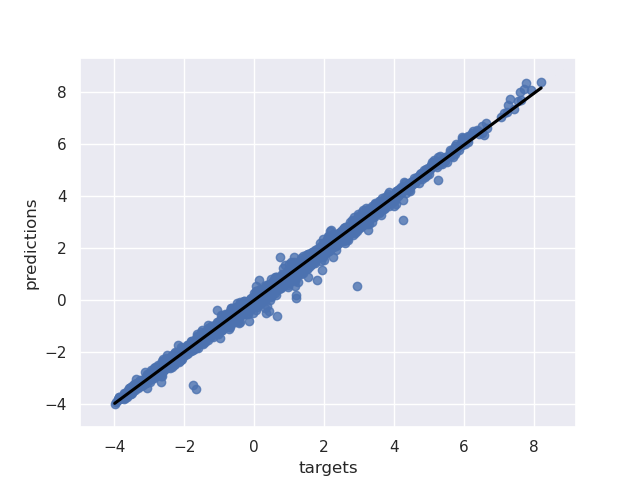}
        \caption{GATGNN (MAE: 0.0727 eV)}
        \vspace{-3pt}
        \label{fig:spGa2D}
    \end{subfigure}
    \begin{subfigure}[t]{0.32\textwidth}
        \includegraphics[width=\textwidth]{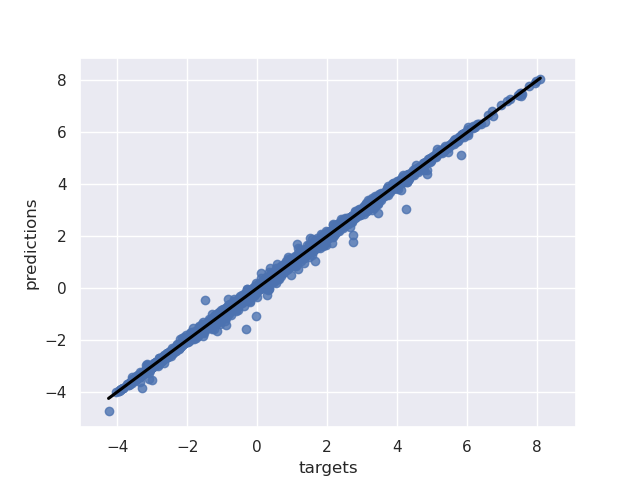}
        \caption{DeeperGATGNN (MAE: 0.0409 eV)}
        \vspace{-3pt}
        \label{fig:dgasp}
    \end{subfigure}
    
    \caption{Comparison of the performance (in 9:1 holdout test) of our DeeperGATGNN  to shallow models over the Alloy Surface Energy dataset. Overall, our model achieves the lowest MAE error for Alloy surface energy prediction.}
    \label{fig:scatterplot}
\end{minipage}
\end{figure}

We also plot the scatter plots of the predicted surface energies versus the true values for the Alloy Surface dataset (Figure~\ref{fig:scatterplot}), over which the DeeperGATGNN achieves the best performance with an MAE score of 0.041 eV. We can see that the plot in Figure~\ref{fig:dgasp} has smallest deviation from the diagonal lines with more narrow distribution of the points.

\FloatBarrier

\subsection{Parameter study of DeeperGATGNN} \label{subsec: parameter}
We perform different parameter studies of our DeeperGATGNN model using 10 graph convolution layers and 500 epochs. We use the Pt-cluster dataset for this purpose. We calculated the results using 5-fold cross validation just as we do in Subsection~\ref{subsec: comparison}.

\begin{figure}[htb!] 
    \begin{subfigure}[t]{0.49\textwidth}
        \includegraphics[width=\textwidth]{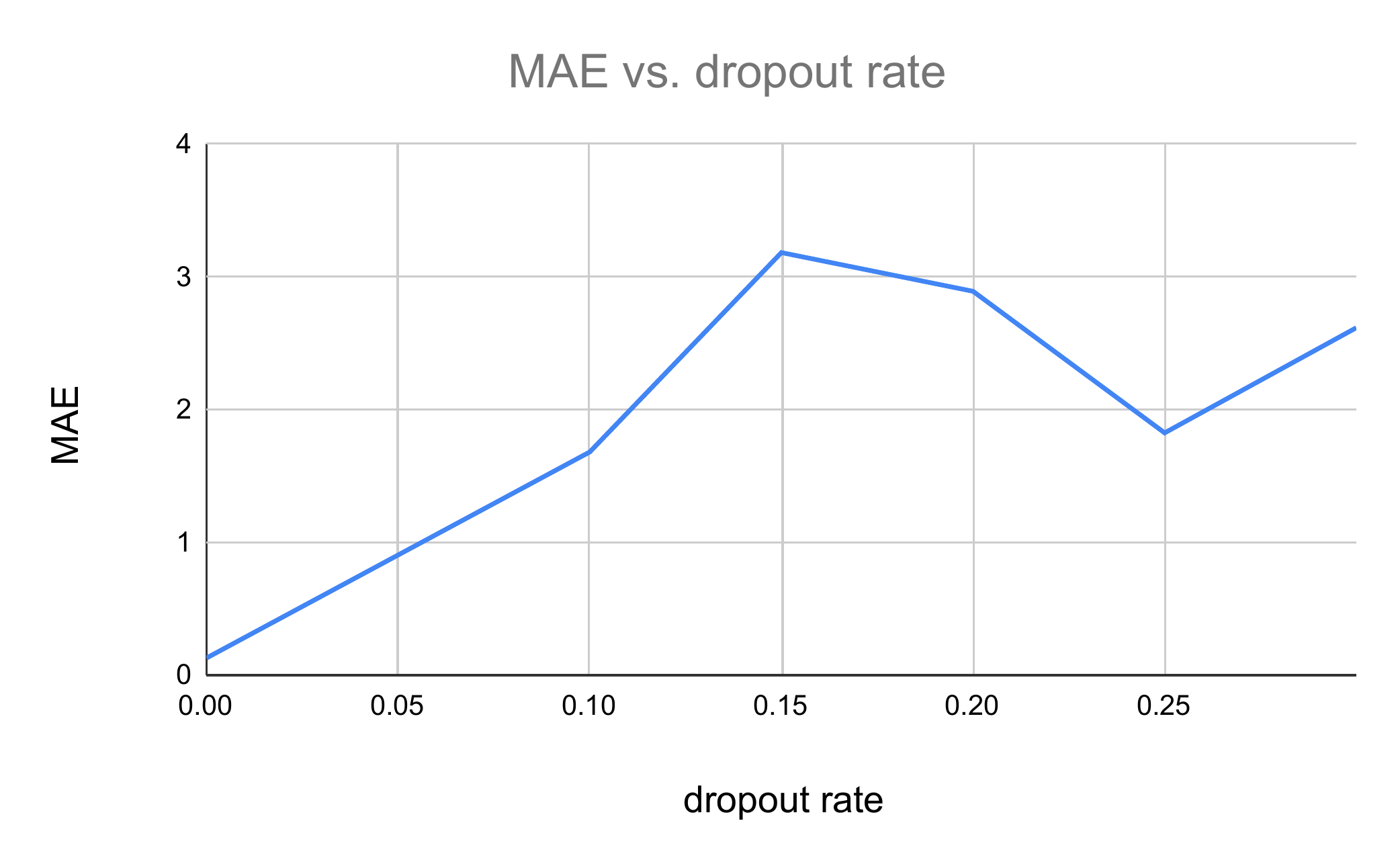}
        \caption{Dropout rate}
        \vspace{-3pt}
        \label{fig:ps-dropout}
    \end{subfigure}\hfill
    \begin{subfigure}[t]{0.49\textwidth}
        \includegraphics[width=\textwidth]{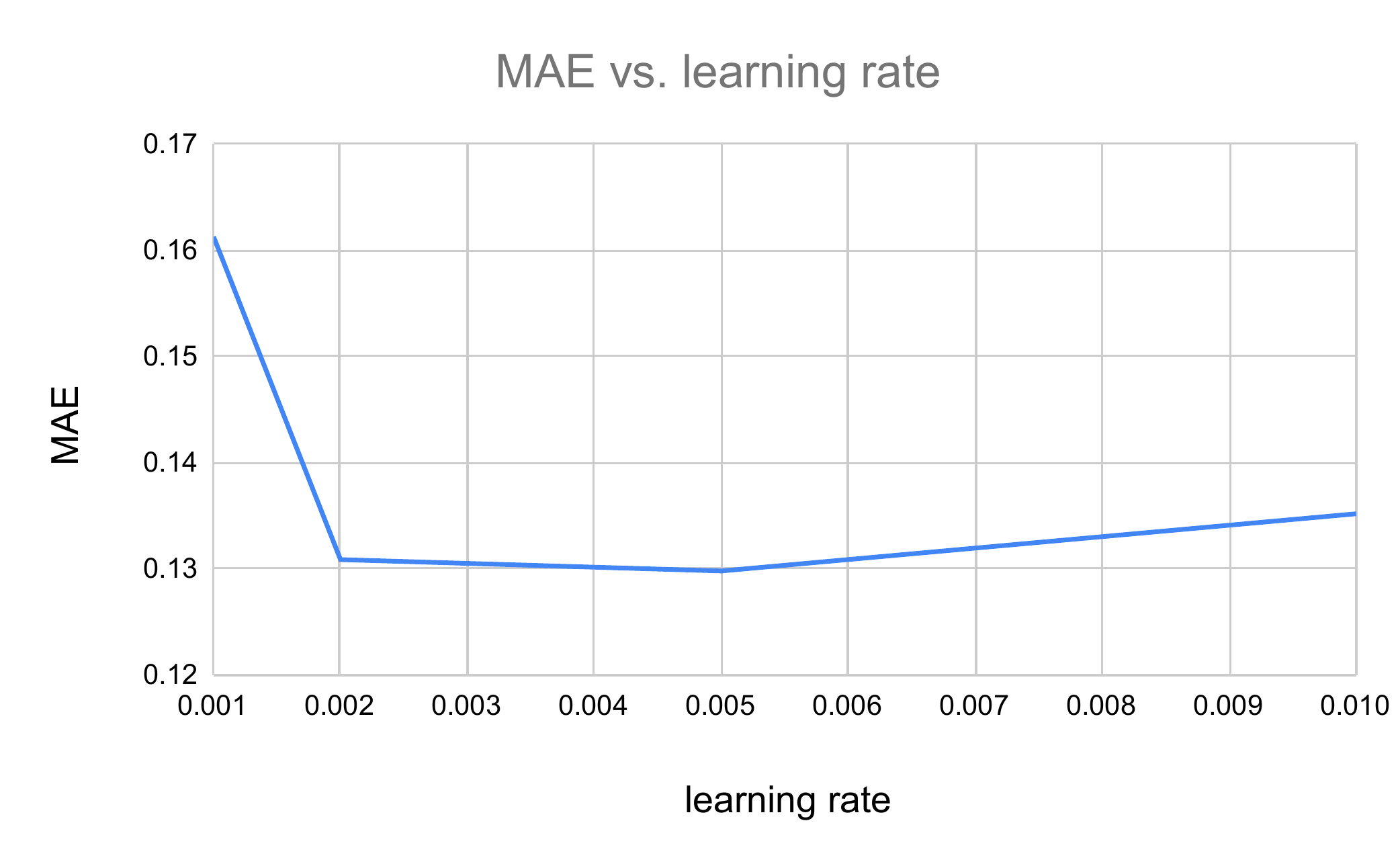}
        \caption{Learning rate }
        \vspace{-3pt}
        \label{fig:ps-lr}
    \end{subfigure}
    \begin{subfigure}[t]{0.49\textwidth}
        \includegraphics[width=\textwidth]{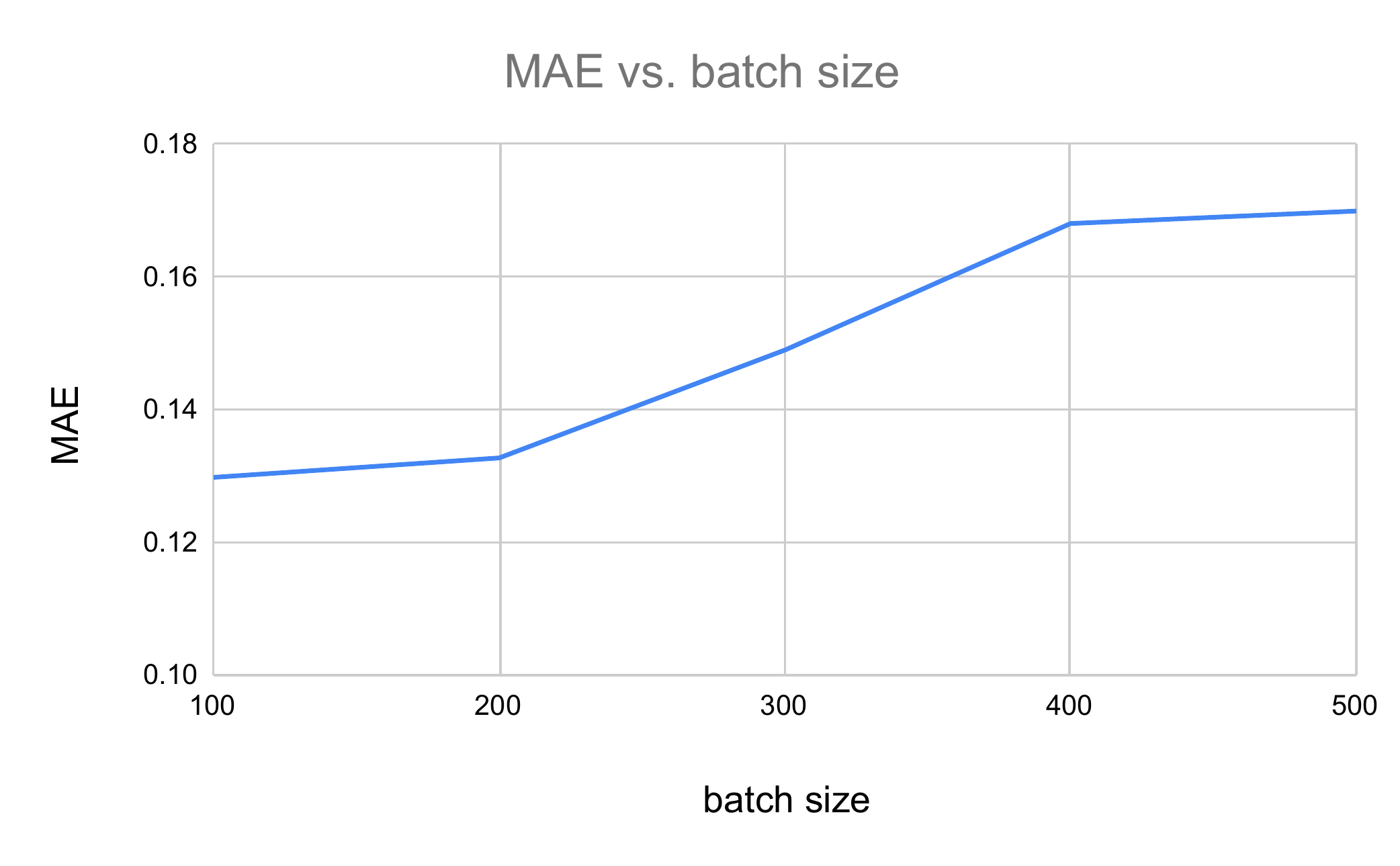}
        \caption{Batch size}
        \vspace{-3pt}
        \label{fig:ps-batch}
    \end{subfigure}\hfill
    \begin{subfigure}[t]{0.49\textwidth}
        \includegraphics[width=\textwidth]{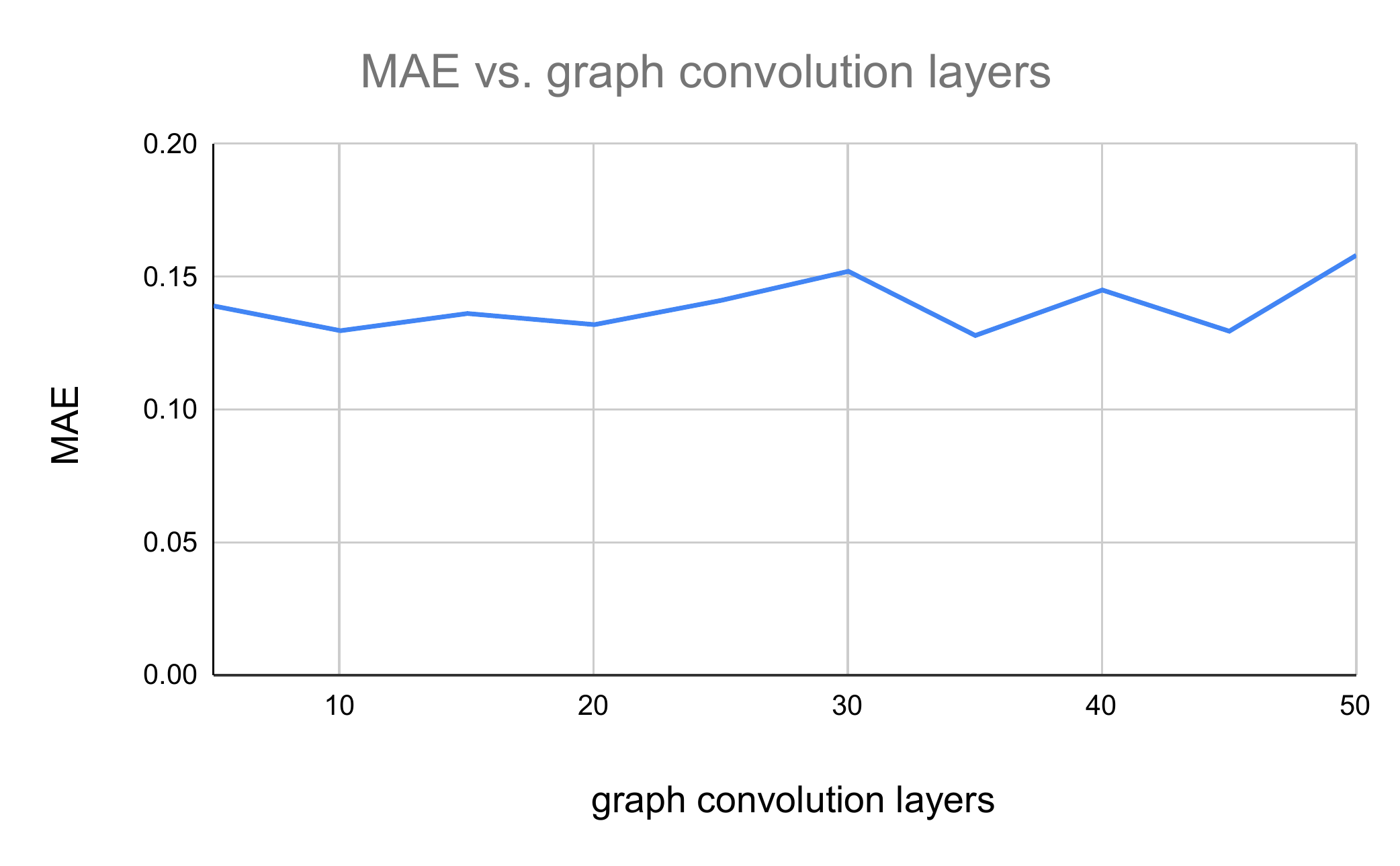}
        \caption{Graph convolution layers}
        \vspace{-3pt}
        \label{fig:ps-gc}
    \end{subfigure}
    
    \caption{Parameter study of DeeperGATGNN - (a) MAE vs dropout rate, (b) MAE vs learning rate, (c) MAE vs batch size (using 10 graph convolution layers) and (d) MAE vs graph convolution layers. All the experiments are done using 500 epochs and 5 fold cross validation on the Pt-cluster dataset.}
    \label{fig:parameterstudy}
\end{figure}

First, we perform experiments on the dropout rate. The best result that we achieved on this dataset in the SOTA performance study is $0.12982$ (MAE) using 10 graph convolution layers and no dropout. We then experiment with varying the dropout rate. We can see the results in Figure~\ref{fig:ps-dropout}. We find that increasing the dropout rate degrades our model's performance. Even with a large number of layers, our model is very scalable and does not need any dropout. Later we will see in this Subsection that our model achieves ever better result using 35 graph convolution layers in which we use no dropout. Also, later we will see in Subsection~\ref{subsec: scalability} that except our model, all the other models perform really poor after a certain number of layers, but our model performs fine even with 30 graph convolution layers and that too without any dropout.

Second, we observe the effect of changing the learning rate in our model. We can see from Figure~\ref{fig:ps-lr} that the best result is achieved for the learning rate value of $0.005$ which we use as the default learning rate for our architecture. We would like to mention that we used a learning rate scheduler library for each of the experiments.

Third, we perform experiments with different batch size with our model. It is experimentally observed that a larger batch sizes usually leads to worse performance than smaller batch sizes~\cite{lecun2012efficient,keskar2016large}. We increased the batch size from 100 (default batch size for our architecture) up to 500 with an interval of 100 and find that our model performs best in the default setting of batch size. We can spot the results of changing batch size in Figure~\ref{fig:ps-batch}.

Next, we investigated the outcome of changing the activation function of our model. By default, our DeeperGATGNN model uses the Softplus activation function. We compared the Softplus activation function result (MAE $0.12982$) with our model with ReLU and Leaky-ReLU activation functions. The MAE values that we obtained for ReLU and Leakey-ReLU are $0.15142$ and $0.15107$ eV, respectively. We can see that the Softplus activation function beats the other activation functions by a good margin.

We then experiment with the performance of our DeeperGATGNN architecture by increasing the number of graph convolution layers with no dropout, and our default batch size, learning rate and activation function. We also use 5 fold cross validation results for this experiment. The impact of graph convolution layers on the DeeperGATGNN performance is shown in Figure\ref{fig:ps-gc}. We can observe that after our previous best result on this dataset (MAE $0.12982$) with 10 graph convolution layers, the performance of our model starts to degrade with the increasing number of layers. It decreases up to 30 layers, then it achieves a new SOTA result (MAE $0.12801$) for this dataset with $35$ graph convolution layers. We examine up to 50 layers and find that this is the best result that we can achieve on this dataset. One important point to notice is that even after 50 layers, the performance of our model has not degraded too much and we achieved SOTA result even after achieving similar result of the 50 layers' for 30 layers. So, there is a possibility that we might achieve even better results if keep going deeper in terms of graph convolution layers in the future on this dataset.

\begin{figure}[H]
\centering
  \includegraphics[width=0.6\linewidth]{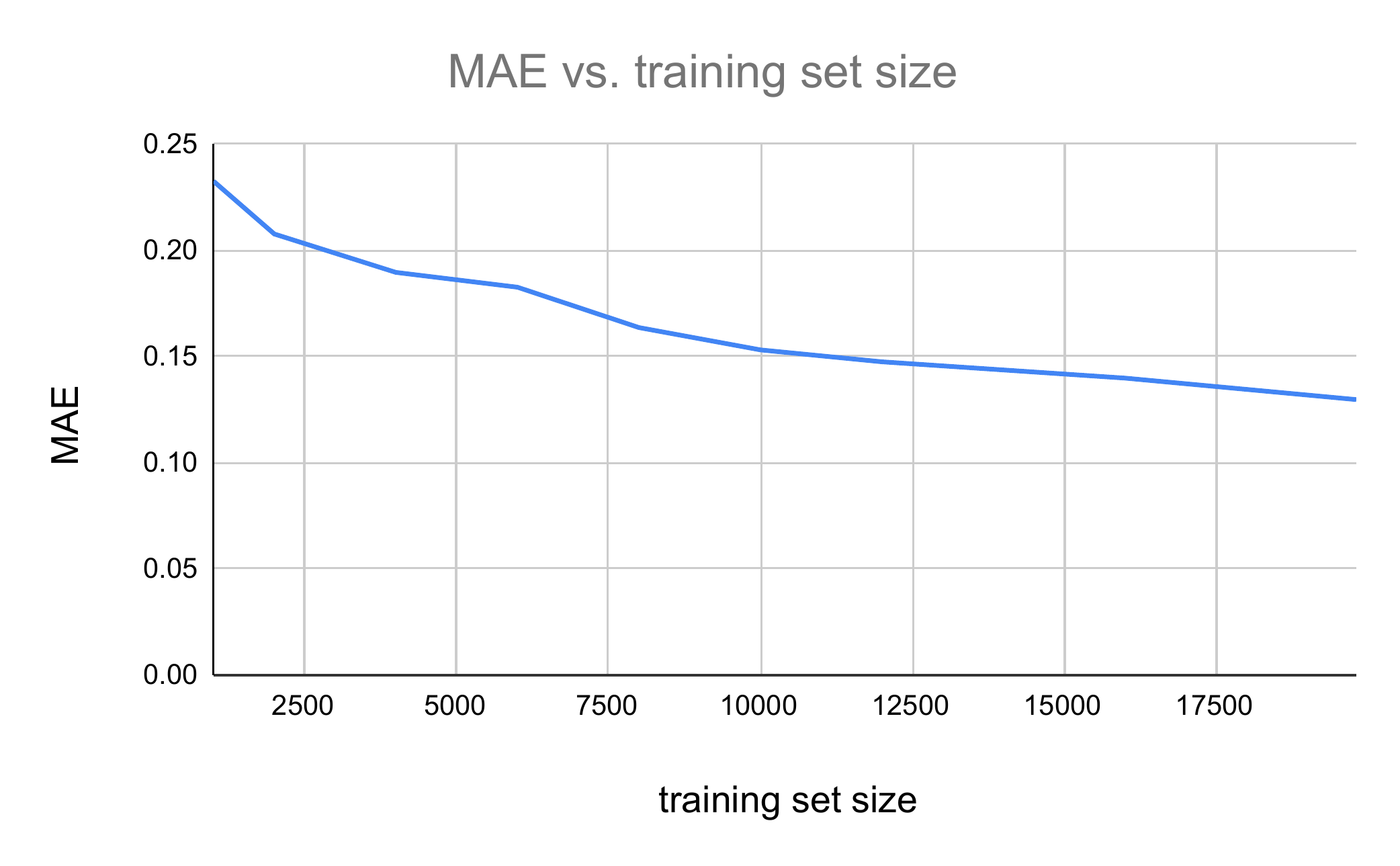}
\caption{Training set size affects the performance of DeeperGATGNN.}
\label{fig:size_dependence}
\end{figure}

Finally, we conduct an experiment to discover the effect of changing the training set size with our model. The success of deep learning architectures largely depends on the amount of training data and our model is no different. We can see from the result which is plotted in Figure~\ref{fig:size_dependence} that the accuracy of our model continues to degrade with decreasing number of training set samples. We believe that if more samples are added to this dataset, our model can achieve even better performance than the SOTA result that we have achieved.

\FloatBarrier

\subsection{Scalability comparison with existing graph neural networks}
\label{subsec: scalability}

We investigate the scalability of our DeeperGATGNN model and other major graph neural networks models heavily used for materials property prediction. We also examine the scalability of other major graph neural networks by adding the differentiable group normalization and skip connections to their architectures. We use the Bulk Formation Energy dataset for this purpose. We conduct all the experiments again using 500 epochs and 5 fold cross validation. We train each of the models for 4, 6, 8, 10, 15, 20, 25 and 30 graph convolution layers and examine their scalability. We limit our experiments to 30 layers because later we show that 30 layers are enough to conclude that our model is the most scalable one among all. We exclude MPNN from this experiment because it has $4385901$ trainable parameters for just 4 graph convolution layers and we also observe from Subsection~\ref{subsec: comparison} that the deeper version of it has about $10929901$ trainable parameters for just 10 layers which is much higher than any other models on average. So the memory it costs and the time it takes to finish training are too large.

\begin{figure}[ht!]
  \centering
  \includegraphics[width=0.9\linewidth]{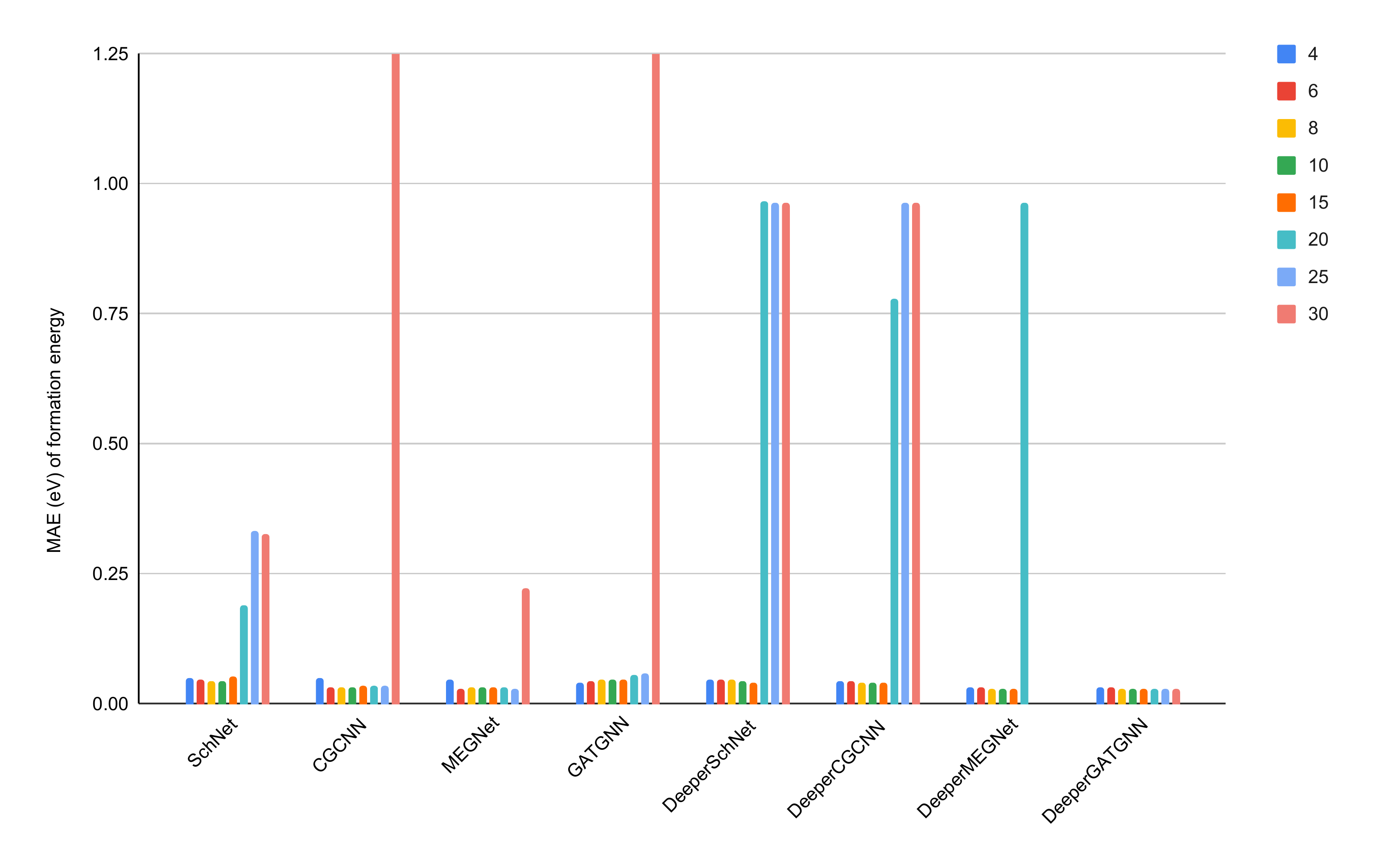}
  \caption{Scalability of graph neural networks in terms of the number of parameters as regard to the depth of the network. Different colors represent different number of graph convolution layers in the networks.}
  \label{fig:scalability}
\end{figure}

First, we examine how the number of parameters changes effect the prediction performance with increasing number of graph convolution layers for the existing graph networks as shown in Figure \ref{fig:scalability}. We can spot that all the existing GNNs crash after a certain number of layers, i.e., the MAE becomes too large so that it can not be used for efficient property prediction any more. For example, SchNet crashes for 20 layers and more. The MAE value of SchNet increases from 0.05158 (15 layers) to 0.18926 (20 layers) where the number of parameters has increased from 1102401 (15 layers) to 1455901 (20 layers).

CGCNN becomes unscalable for 30 layers. The MAE value and number of trainable parameters increase from 0.0351 and 1337101, respectively, (25 layers) to 2.27757 and 1590101, respectively (30 layers). The same goes for MEGNet as well, it also crashes for 30 layers and the MAE value and number of parameters increase from 0.03024 and 4857201, respectively, (25 layers) to 0.22373 and 5819201, respectively (30 layers). But the change of MAE value is not as drastic as that of CGCNN although MEGNet has much higher number of parameters than that of CGCNN. For better visualization, we limit the y-axis of Figure~\ref{fig:scalability} to 1.25, that is why the plotting of MAE value gets trimmed for CGCNN for 30 layers.

GATGNN also crashes for 30 layers. It also has a drastic change of performance from 25 to 30 layers like CGCNN. The MAE and the number of parameters rise from 0.05751 and 863370, respectively, (25 layers) to 2.7937 and 1030770, respectively (30 layers). We can see that both GATGNN and CGCNN have much smaller number of trainable parameters compared to that of MEGNet and they are also very similar in terms of scalability.

As we see that, none of the existing GNNs are scalable for 30 layers and more, so next we examine these model's scalability by adding differentiable group normalization and skip connections (deeper versions). DeeperSchNet again crashes for 20 layers which means the modification did not affect the performance. The MAE value and the number of parameters change from 0.04209 and 1144401, respectively, (15 layers) to 0.96561 and 1511901, respectively (20 layers). Although the result for 15 layers of DeeperSchNet is better than that of the original SchNet, surprisingly the change of MAE for 20 layers is much more drastic in DeeperSchNet than that of SchNet.

Both DeeperCGCNN and DeeperMEGNet perform worse than the original CGCNN and MEGNet, respectively. DeeperCGCNN becomes unscalable for 20 layers which is quicker than original SchNet (30 layers). The MAE and number of parameters increase from 0.04253 and 873101, respectively, (15 layers) to 0.7783 and 1140101, respectively (20 layers). DeeperMEGNet also becomes unscalable for 20 layers. The MAE and number of parameters increase from 0.03 and 3185201, respectively, (15 layers) to 0.96507 and 4231201, respectively (20 layers). We did not do experiments with DeeperMEGNet for 25 and 30 layers because it already became unscalable and the number of parameters was already huge for just 20 layers.

Now we discuss about the scalability of our DeeperGATGNN model. We can see that our model achieves the SOTA result for 15 layers on this dataset. Our model is the only model that did not crash for even 30 layers. Also the number of trainable parameters of our model is the smallest (except GATGNN) among other GNNs for each layer. The MAE and number of parameters for 30 layers of our model is 0.03041 and 1089906, respectively. One of the key points to notice here is that, though our model is scalable even up to 30 layers, the performance did not improve after 15 layers. But we can see from Subsection~\ref{subsec: parameter} that performance on the Pt-cluster dataset also did not improve after 10 layers, but it did not crash like this experiment here and finally it was able to improve for 35 layers. So as our model is still scalable for 30 layers, there is a strong possibility that our model might achieve even better result on the bulk formation energy dataset if we go deeper in terms of graph convolution layers. But 30 layers was enough to show that our DeeperGATGNN model is the most scalable model among all. We would like to mention that, we also tried using a little dropout with 20, 25 and 30 layers with our model but it did not improve our result.

\FloatBarrier

\subsection{Physical insights from scalable DeeperGATGNN}
We dig into some physical insights of our DeeperGATGNN architecture. We use the t-distributed stochastic neighbor embedding (t-SNE)~\cite{van2008visualizing} for this purpose which is a very well used non-linear technique for visualizing high dimensional data. The objective of t-SNE is to map higher dimensional data points to very low dimensional points (usually 2d or 3d). The pairwise distance between data points are well preserved after mapping~\cite{van2008visualizing,li2017application}. So closer points in higher dimension tend to remain closer after mapping to the lower dimension.

\begin{figure}[H] 
    \begin{subfigure}[t]{0.49\textwidth}
        \includegraphics[width=\textwidth]{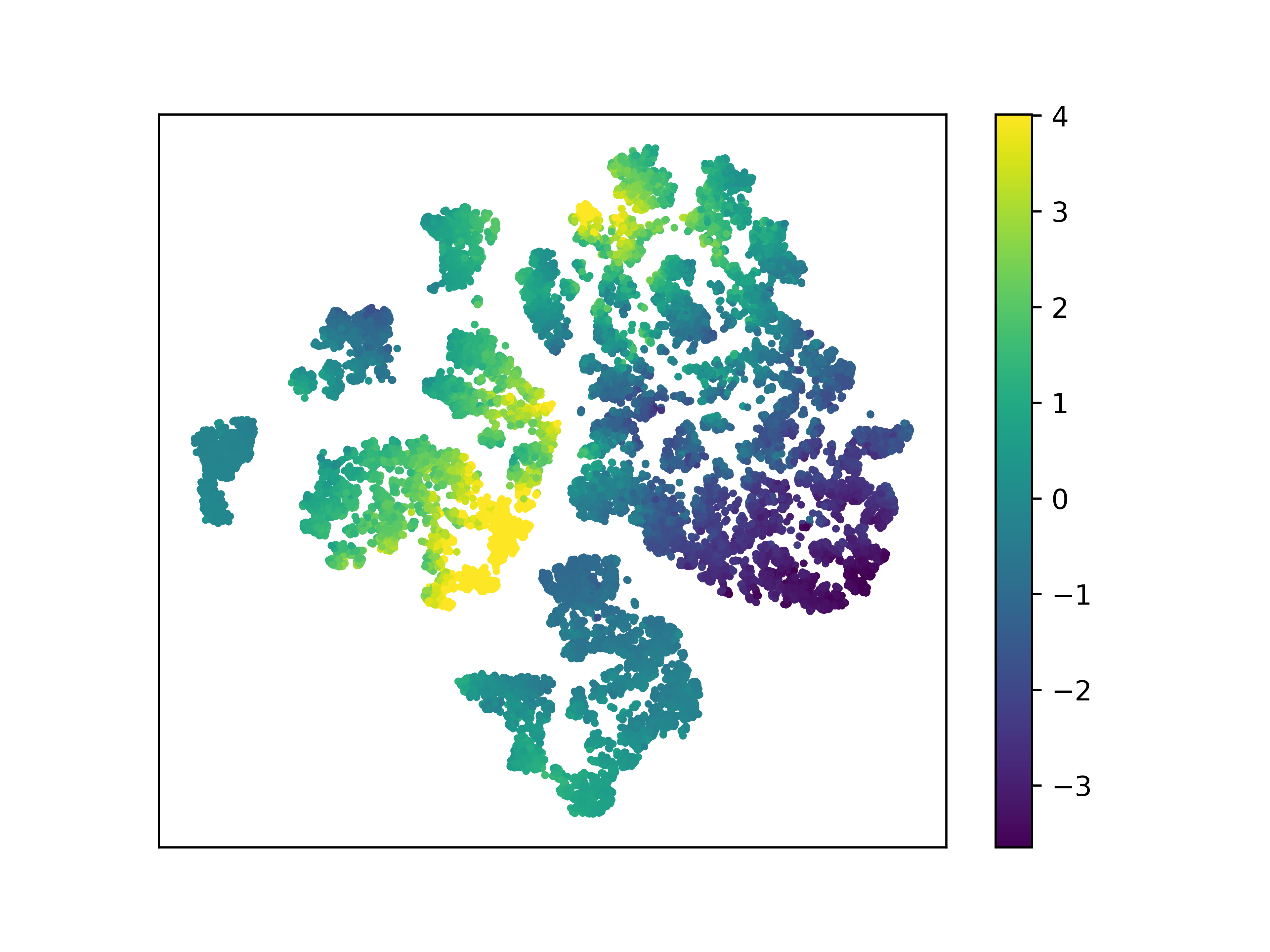}
        \caption{CGCNN}
        \vspace{-3pt}
        \label{fig:resnet}
    \end{subfigure}\hfill
    \begin{subfigure}[t]{0.49\textwidth}
        \includegraphics[width=\textwidth]{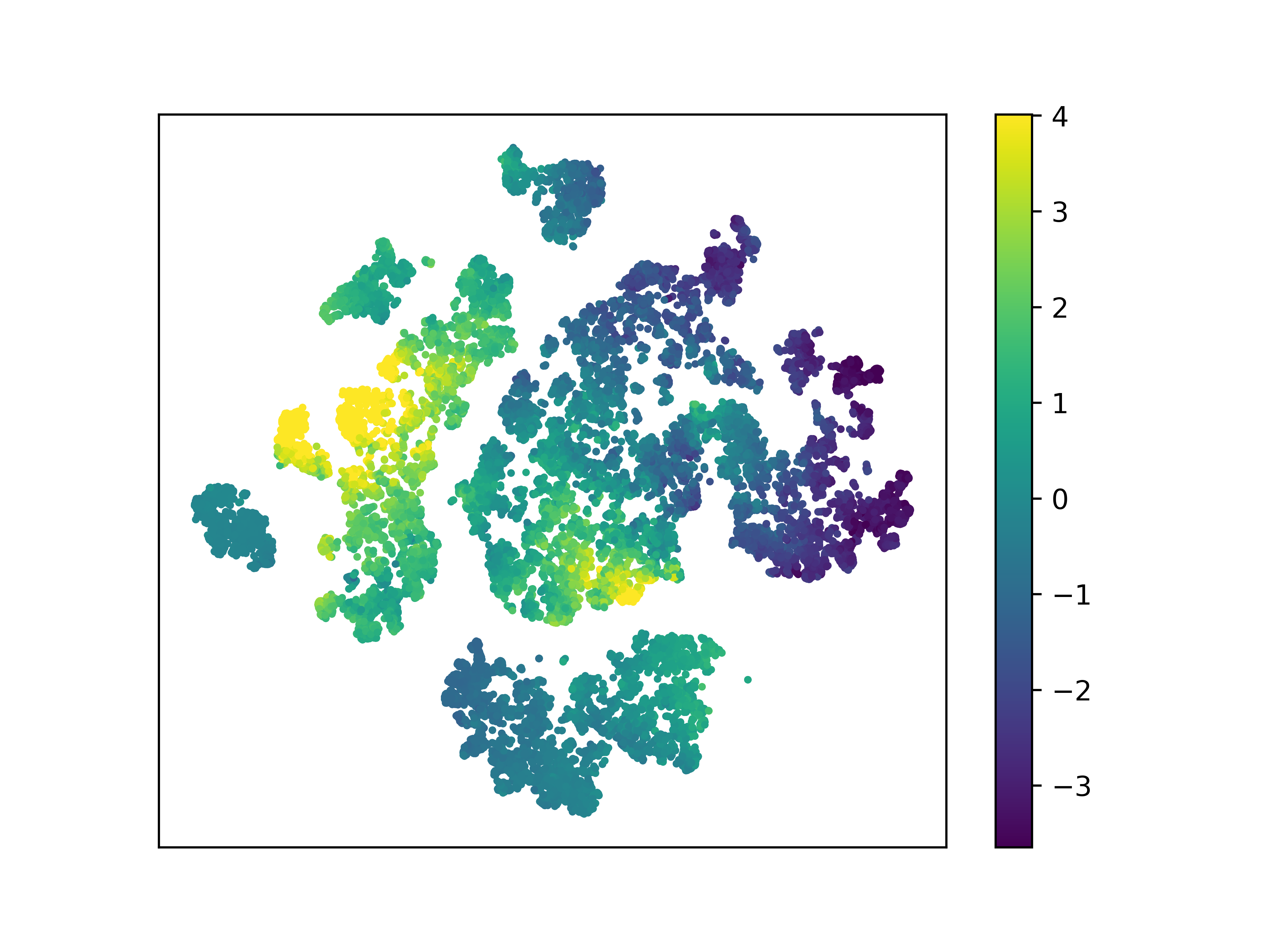}
        \caption{MEGNet}
        \vspace{-3pt}
        \label{fig:La3Al1N1}
    \end{subfigure}
    
    \begin{subfigure}[t]{0.49\textwidth}
        \includegraphics[width=\textwidth]{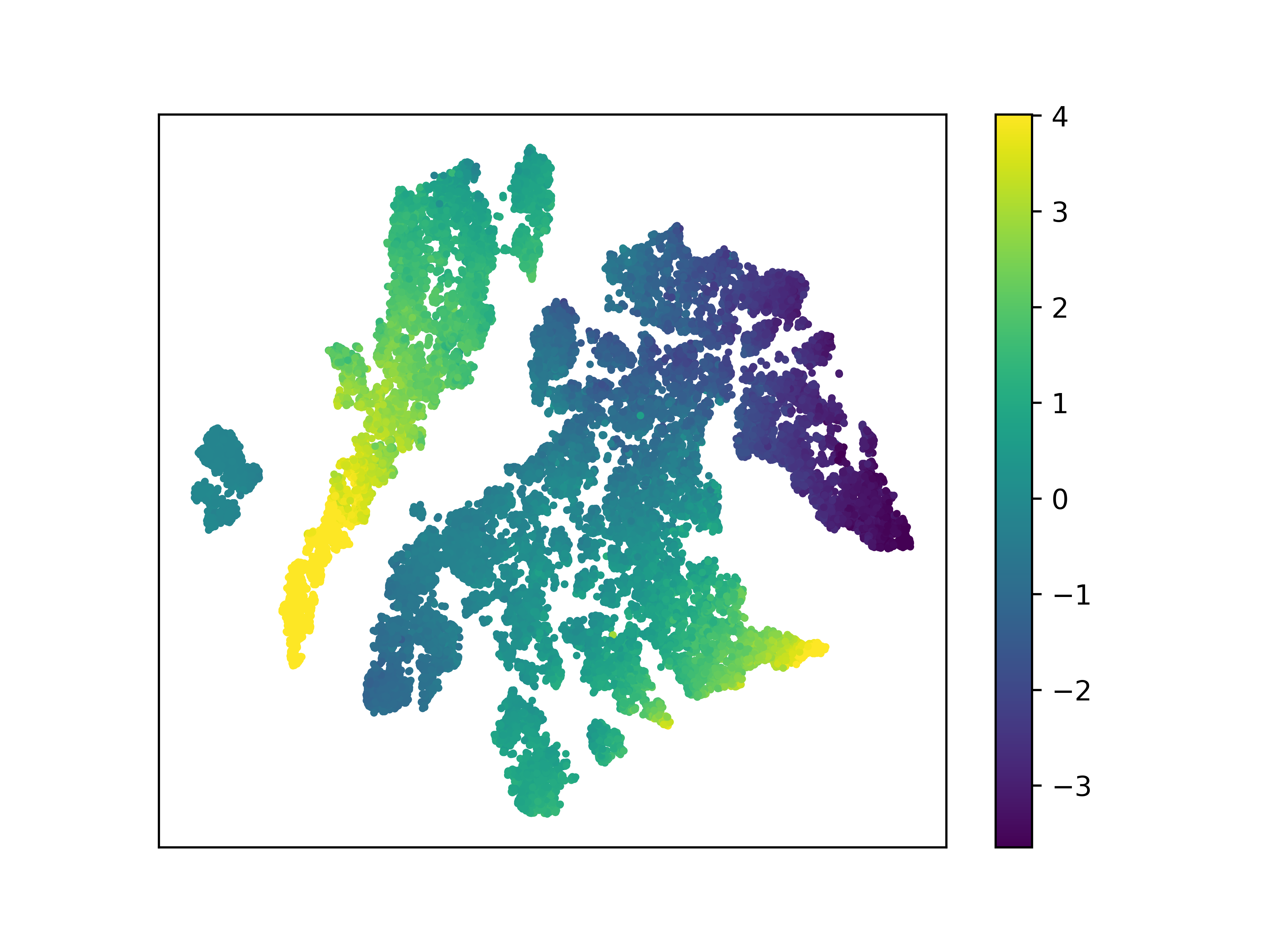}
        \caption{GATGNN}
        \vspace{-3pt}
        \label{fig:resnet}
    \end{subfigure}\hfill
    \begin{subfigure}[t]{0.49\textwidth}
        \includegraphics[width=\textwidth]{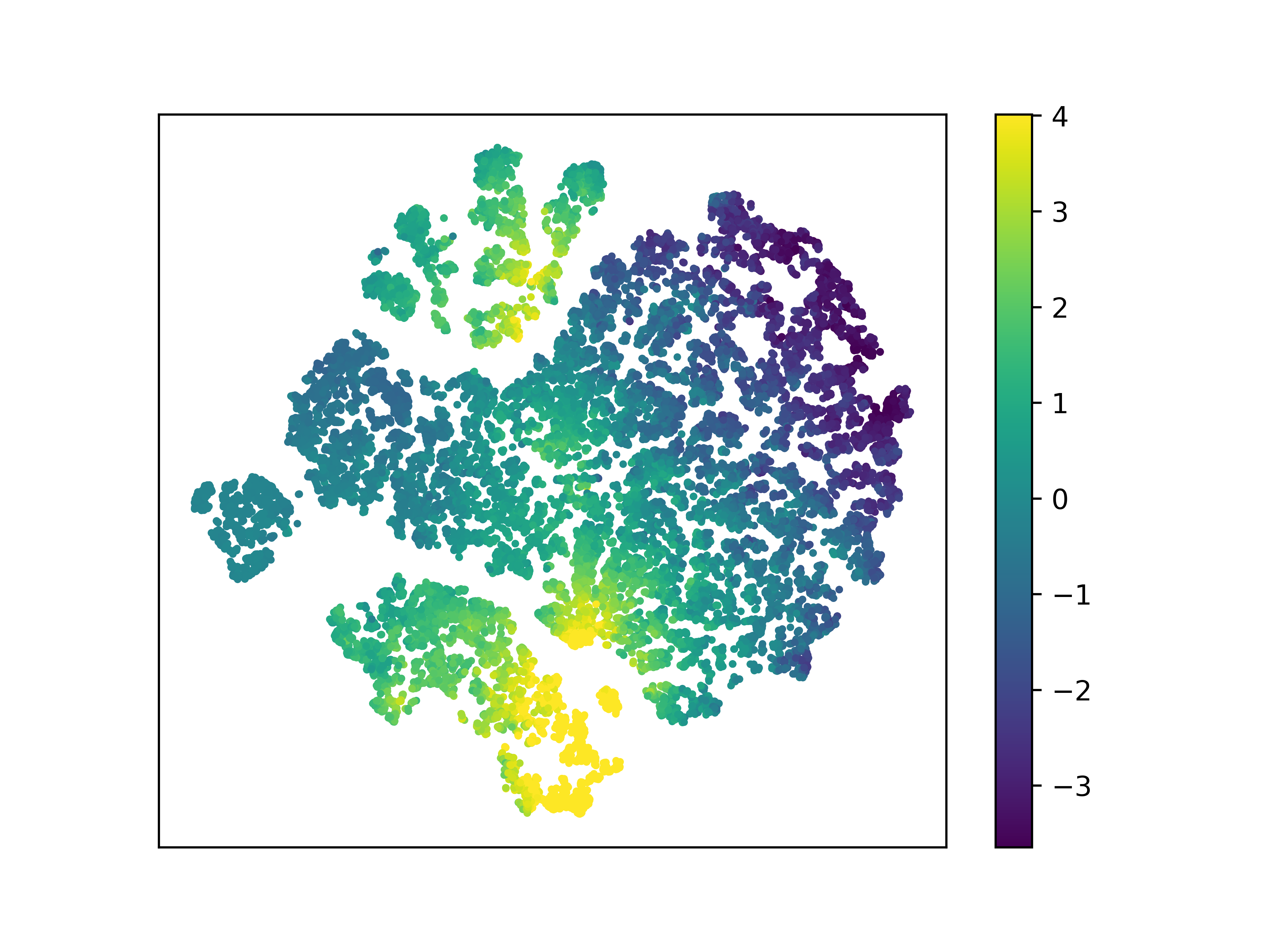}
        \caption{DeeperGATGNN}
        \vspace{-3pt}
        \label{fig:La3Al1N1}
    \end{subfigure}
    \caption{Distribution of latent space}
    \label{fig:tsne}
\end{figure}

Here, we use the Alloy Surface dataset for visualizing our model, as well as, GATGNN, CGCNN and MEGNet models. We fetch the output of the first layer after the final graph convolution layer for the t-SNE plots which can be viewed in Figure~\ref{fig:tsne}. Different colors represent different alloy surface energy in the latent space. We can see from Figure~\ref{fig:tsne} that different groups are formed after the 2-dimensional mapping and materials (points) in the same cluster have a very good probability of having similarities in their composition and structure. Each model might generate different latent space, but we can still get a good idea about the pattern of their prediction by analyzing these clusters. We can see that some area in all the images have been colored with the same color implying that they have got some similar patterns. Although it is important to mention that sizes of clusters and the distance between them do not bear much significance in a t-SNE plot~\cite{wattenberg2016use}.

\section{Discussion}
\label{sec:others}

Graph neural networks are increasingly used for solving challenging problems in materials and physics \cite{shlomi2020graph,sanchez2020learning,park2020developing,park2021accurate}. Compared to other representations, representations in GNNs are inherently rotation and translation invariant, making it ideal to model the atomic relationships. However, we find that there are several key issues in designing and training scalable GNN models. 

The first pitfall is the GNN models can easily go under-trained due to the high computational complexity in training complex GNNs with from hundreds of thousands to several million parameters. The situation becomes even worse when one has to do large-scale hyper-parameter tuning. For example, in the benchmark mark study of GNNs, the hyper-parameters include three encoding dimensions, convolution layer number, fully connected layer number, pooling methods, the learning rate and the batch size. And the study found that existing GNNs tend to achieve optimal performance for different datasets using very different hyper-parameter sets. This expensive hyper-parameter search process forces them to use only 200 epochs for evaluation. However, our analysis in Figure \ref{fig: 3} shows that their networks are all under-trained, which won't stagnate until 500 epochs. This has led to their severe under-estimation of the GNN performances for all the results they reported. For example for the Alloy dataset the CGCNN MAE with 250 epochs of training is 0.06 eV  40\% larger than the result (0.042 eV) when using 500 epochs of training  Since running more epochs with huge hyper-parameter space is prohibitive or infeasible, it is then more desirable to use GNN models that can achieve more stable results with default or minor parameter tuning. For example, our deeperGATGNN models achieved the state-of-the-art results across five datasets using the same architecture except with varying number of convolution layers. 

The second pitfall for GNNs is that they usually suffer from the over-smoothing issue which lead to their performance degradation when too many layers are used \cite{li2018deeper}. This is clearly shown in our scalability study in Section \ref{sec:xxx} and Figure \ref{fig:scalability}. All existing GNN algorithms except our DeeperGATGNN have significant performance degradation when 30 convolution layers are used. It is interesting to find that while differentiable normalization and skip connection have effectively helped our DeeperGATGNN to address this issue, the same strategy does not work equally well for ScheNet, CGCNN, and MEGNet even though it does help to improve their performance too.

Another limit to our model performance is the availability of data or the information input to our models. For example in the 2D materials dataset which has only 3814 samples, MEGNet achieved the best result with MAE of 0.17194 with 816,801 parameters. Our DeeperGATGNN with 25 convolution layers and 913,546 parameters achieves an MAE of 0.17185. In this case, it seems that to further improve the performance, additional information such as the angular information of the structures of materials are needed.

\section{Conclusion}
\label{sec:others}

Large scale very deep neural networks have generated breakthrough results in a variety of application domains except the materials property prediction. Existing graph neural networks for materials property prediction have so far all suffered from the over-smoothing issue and cannot scale up to very deep networks without significant performance degradation. Here we proposed a scalable global attention graph neural network DeeperGATGNN for achieving state-of-the-art materials property prediction with up to 10\% performance improvement over the best results of all previous graph neural networks for five out of the six benchmark datasets. This is all achieved with a single neural architecture and hyper-parameter set except for an easy-to-set large-enough graph convolution layer. This makes it much simpler in practical materials property prediction without the need for expensive hyper-parameter tuning. Our deeper graph neural network enabling strategies such as the ResNet skip connection and differentiable group normalization have shown to be able to also improve the scalability and performance of other existing graph neural networks such as MEGNet and SchNet and CGCNN, but only on special datasets.

\section{Availability of data}

The data that support the findings of this study are openly available in Materials Project database at \href{http:\\www.materialsproject.org}{\textcolor{blue}{http:\\www.materialsproject.org}}

\section{Contribution}
Conceptualization, J.H.; methodology, J.H., S.O., and S.L.; software, S.O. and S.L. ; validation, S.O. and J.H.;  investigation, J.H., S.O., S.L., N.F., L.W., S.D., R.D., Q.L; resources, J.H.; data curation, J.H., and S.O.; writing--original draft preparation,J.H., S.O, S.L.,L.W. ; writing--review and editing, J.H, S.O., N.F.,S.D.,R.D.,Q.L.; visualization, J.H., S.O., S.L.; supervision, J.H.;  funding acquisition, J.H.

\section{Acknowledgement}
Research reported in this work was supported in part by NSF under grants 1940099 and 1905775. The views, perspective, and content do not necessarily represent the official views of NSF. We would like to thank Daniel Varivoda for proof-reading the paper.

\bibliographystyle{unsrt}  
\bibliography{references}  
\end{document}